\documentclass[10pt,aps,amsmath,pre,preprint,groupedaddress,showpacs]{revtex4}
\usepackage{graphicx}
\usepackage{dcolumn}
\usepackage{bm}
\begin{document}

\title{Synchronization in disordered Josephson junction arrays:\\ Small-world connections and the Kuramoto model}

\author{B. R. Trees}
  \email{brtrees@owu.edu}
\author{V. Saranathan}
\affiliation{Department of Physics and Astronomy \\ Ohio Wesleyan University}
\author{D. Stroud}
\affiliation{Department of Physics \\The Ohio State University}

\date{\today}

\begin{abstract}
We study synchronization in disordered arrays of Josephson junctions. In the first half of the paper, we consider the relation between the coupled resistively- and capacitively shunted junction (RCSJ) equations for such arrays and effective phase models of the Winfree type. We describe a multiple-time scale analysis of the RCSJ equations for a ladder array of junctions \textit{with non-negligible capacitance} in which we arrive at a second order phase model that captures well the synchronization physics of the RCSJ equations for that geometry. In the second half of the paper, motivated by recent work on small world networks, we study the effect on synchronization of random, long-range connections between pairs of junctions. We consider the effects of such shortcuts on ladder arrays, finding that the shortcuts make it easier for the array of junctions in the nonzero voltage state to synchronize. In 2D arrays we find that the additional shortcut junctions are only marginally effective at inducing synchronization of the active junctions. The differences in the effects of shortcut junctions in 1D and 2D can be partly understood in terms of an effective phase model.
\end{abstract}

\pacs{05.45.Xt,05.45.-a,74.50.+r}

\maketitle

\section{\label{sec:intro} Introduction}

The synchronization of coupled nonlinear oscillators has been a fertile area of research for decades\cite{Pikovsky01}. In particular, phase models of the Winfree type\cite{Winfree67} have been extensively studied. In one dimension, a generic version of this model for $N$ oscillators is
\begin{equation}
\frac{d\theta_j}{dt}=\Omega_j + \sum_{k=1}^{N}\sigma_{j,k}\Gamma\left(\theta_k-\theta_j\right),
\label{eq:winfree}
\end{equation}
where $\theta_j$ is the phase of oscillator $j$, which can be envisioned as a point moving around the unit circle with angular velocity $d\theta_j/dt$. In the absence of coupling, this overdamped oscillator has an angular velocity $\Omega_j$. $\Gamma(\theta_k-\theta_j)$ is the coupling function, and $\sigma_{j,k}$ describes the range and nature ({\em e.g.} attractive or repulsive) of the coupling. The special case $\Gamma(\theta_k -\theta_j)=\sin(\theta_k -\theta_j)$, $\sigma_{j,k}=\alpha/N$ ($\alpha=$ constant), corresponds to the uniform, sinusoidal coupling of each oscillator to the remaining $N-1$ oscillators. This mean-field system is usually called the (globally-coupled) Kuramoto model (GKM). Kuramoto was the first to show that for this particular form of coupling and in the $N\rightarrow\infty$ limit, there is a continuous dynamical phase transition at a critical value of the coupling strength $\alpha_c$ and that for $\alpha >\alpha_c$ both phase and frequency synchronization appear in the system\cite{Kuramoto84,Strogatz2000}. If $\sigma_{j,k}=\alpha\delta_{j,k\pm 1}$ while the coupling function retains the form $\Gamma(\theta_j - \theta_k)=\sin(\theta_k -\theta_j)$ we have the so-called locally-coupled Kuramoto model (LKM), in which each oscillator is coupled only to its nearest neighbors. Studies of synchronization in the LKM\cite{Sakaguchi87}, including extensions to more than one spatial dimension, have shown that $\alpha_c$ grows without bound in the $N\rightarrow\infty$ limit\cite{Strogatz88}.

Several years ago, Watts and Strogatz introduced a simple model for tuning collections of coupled dynamical systems between the two extremes of random and regular networks\cite{Watts98}. In this model, connections between nodes in a regular array are randomly rewired with a probability $p$, such that $p=0$ means the network is regularly connected, while $p=1$ results in a random connection of nodes. For a range of intermediate values of $p$ between these two extremes, the network retains a property of regular networks (a large clustering coefficient) and also acquires a property of random networks (a short characteristic path length between nodes). Networks in this intermediate configuration are termed ``small-world'' networks. Many examples of such small worlds, both natural and human-made, have been discussed\cite{Strogatz2001}. Not surprisingly, there has been much interest in the synchronization of dynamical systems connected in a small-world geometry\cite{Barahona2002,Nishikawa2003}. Generically, such studies have shown that the presence of small-world connections make it easier for a network to synchronize, an effect generally attributed to the reduced path length between the linked systems. This has also been found to be true for the special case in which the dynamics of each oscillator is described by a Kuramoto model\cite{Hong2002a,Hong2002b}.

As an example of physically-controllable systems of nonlinear oscillators which can be studied both theoretically and experimentally, Josephson junction (JJ) arrays are almost without peer. Through modern fabrication techniques and careful experimental methods one can attain a high degree of control over the dynamics of a JJ array, and many detailed aspects of array behavior have been studied\cite{Newrock2000}. Among the many different geometries of JJ arrays, \textit{ladder} arrays (see Fig.~\ref{fig:ladder}) deserve special attention. For example, they have been observed to support stable time-dependent, spatially-localized states known as discrete breathers\cite{Fistul2000}. In addition, the ladder geometry is more complex than that of better understood serial arrays but less so than fully two-dimensional (2D) arrays. In fact, a ladder can be considered as a special kind of 2D array, and so the study of ladders could throw some light on the behavior of such 2D arrays. Also, linearly-stable synchronization of the horizontal, or rung, junctions in a ladder (see Fig.~\ref{fig:ladder}) is observed in the absence of a load over a wide range of dc bias currents and junction parameters (such as junction capacitance), so that synchronization in this geometry appears to be robust\cite{Trees2001a}.

In the mid 1990's it was shown that a \textit{serial} array of zero-capacitance, {\it i.e.} overdamped, junctions coupled to a load could be mapped onto the GKM\cite{Wiesenfeld96,Wiesenfeld98}. The load in this case was essential in providing an all-to-all coupling among the junctions. The result was based on an averaging process, in which (at least) two distinct time scales were identified: the ``short'' time scale set by the rapid voltage oscillations of the junctions (the array was current biased above its critical current) and ``long'' time scale over which the junctions synchronize their voltages. If the resistively-shunted junction (RSJ) equations describing the dynamics of the junctions are integrated over one cycle of the ``short'' time scale, what remains is the ``slow'' dynamics, describing the synchronization of the array. This mapping is useful because it allows knowledge about the GKM to be applied to understanding the dynamics of the serial JJ array. For example, the authors of Ref.~\cite{Wiesenfeld96} were able, based on the GKM, to predict the level of critical current disorder the array could tolerate before frequency synchronization would be lost. Frequency synchronization, also described as entrainment, refers to the state of the array in which all junctions \textit{not} in the zero-voltage state have equal (to within some numerical precision) time-averaged voltages: $(\hbar/2e)\langle d\theta_j/dt\rangle_t$, where $\theta_j$ is the gauge-invariant phase difference across junction $j$.  More recently, the ``slow'' synchronization dynamics of finite-capacitance serial arrays of JJ's has also been studied\cite{Chernikov95,Watanabe97}. Perhaps surprisingly, however, no experimental work on JJ arrays has verified the accuracy of this GKM mapping. Instead, the first detailed experimental verification of Kuramoto's theory was recently performed on systems of coupled electrochemical oscillators\cite{Kiss2002}. 

Recently, Daniels {\text et al.}\cite{Daniels2003}, with an eye toward a better understanding of synchronization in 2D JJ arrays, showed that a ladder array of \textit {overdamped} junctions could be mapped onto the locally-coupled Kuramoto model (LKM). This work was based on an averaging process, as in Ref.~\cite{Wiesenfeld96}, and was valid in the limits of weak critical current disorder (less than about $10\%$) and large dc bias currents, $I_B$, along the rung junctions ($I_B/\langle I_{c}\rangle \agt 3$, where $\langle I_{c}\rangle$ is the arithmetic average of the critical currents of the rung junctions. The result demonstrated, for both open and periodic boundary conditions, that synchronization of the current-biased rung junctions in the ladder is well described by Eq.~\ref{eq:winfree}.
 
The goal of the present work is twofold. First, we will demonstrate that a ladder array of \textit{underdamped} junctions can be mapped onto a second-order Winfree-type oscillator model of the form
\begin{equation}
a\frac{d^2\theta_j}{dt^2} + \frac{d\theta_j}{dt} = \Omega_j + \sum_{k=1}^{N}\sigma_{j,k}\Gamma(\theta_k - \theta_j),
\label{eq:winfree2}
\end{equation}
where $a$ is a constant related to the average capacitance of the rung junctions. This result is based on the resistively and capacitively-shunted junction (RCSJ) model and a multiple time scale analysis of the classical equations for the array. Secondly, we study the effects of small world (SW) connections on the synchronization of both overdamped and underdamped ladder arrays. We will demonstrate that SW connections make it easier for the ladder to synchronize, and that a Kuramoto or Winfree type model (Eqs.~\ref{eq:winfree} and~\ref{eq:winfree2}), suitably generalized to include the new connections, accurately describes the synchronization of this ladder.

This paper is organized as follows. In Secs.~\ref{sec:twotiming} and~\ref{sec:results} we discuss the multiple time-scale technique for deriving the coupled phase oscillator model for the \textit{underdamped} ladder \textit{without} SW connections. We compare the synchronization of this ``averaged'' model to the exact RCSJ behavior. We also analyze how the array's synchronization depends on the capacitance of the junctions. In Sec.~\ref{sec:SW}, we study the effects of SW connections, or shortcuts, on the synchronization of both overdamped and underdamped ladders. In our scenario, each SW connection is actually another Josephson junction. We generalize our phase-oscillator model to include the effects of shortcuts and relate our results to earlier work on Kuramoto-like models in the presence of shortcuts\cite{Hong2002a,Hong2002b}. In Sec.~\ref{sec:2D} we study the effects of SW connections on synchronization in disordered 2D arrays. Here we find that the disordered 2D array, which does \textit{not} fully synchronize in the pristine case (\textit{i.e.} in the absence of shortcuts), is only weakly synchronized by the addition of shortcut junctions between superconducting islands in the array.  In Sec.~\ref{sec:conclusion} we conclude and discuss possible avenues for future work. 

\section{\label{sec:twotiming}Phase Model for Underdamped Ladder}

\subsection{\label{sec:background} Background}

\begin{figure}
\includegraphics[scale=0.60]{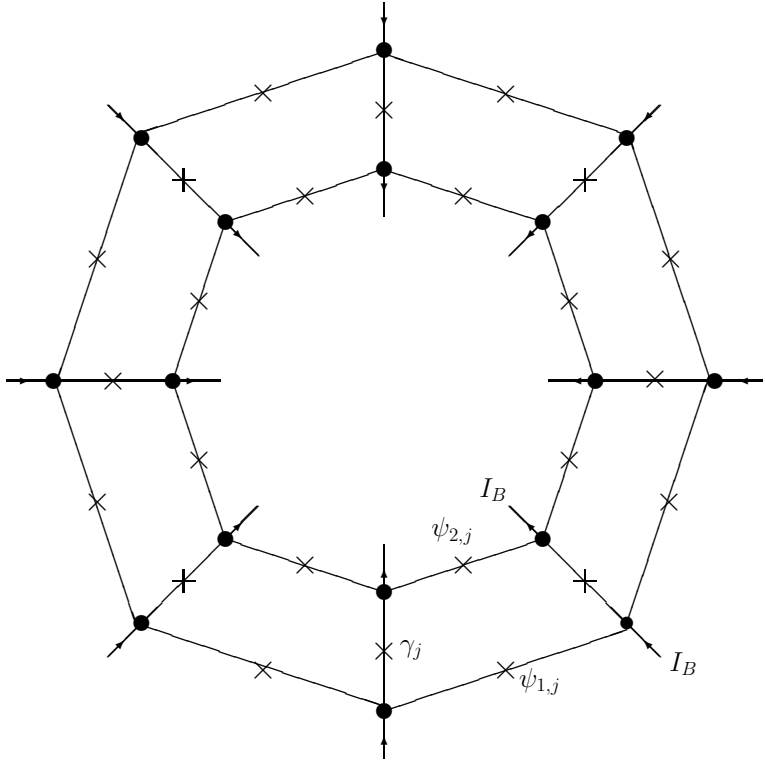}
\caption{\label{fig:ladder} Ladder array with periodic boundary conditions and $N=8$ plaquettes. A uniform, dc bias current $I_B$ is inserted into and extracted from each rung as shown. The gauge-invariant phase difference across the rung junctions is denoted by $\gamma_j$ where $1\leq j\leq N$, while the corresponding quantities for the off-rung junctions along the outer(inner) edge are $\psi_{1,j}$($\psi_{2,j}$). The rung junctions are assumed to be disordered while the off-rung junctions are uniform.}
\end{figure}

The ladder geometry is shown in Fig.~\ref{fig:ladder}, which depicts an array with $N=8$ plaquettes, periodic boundary conditions, and uniform dc bias currents, $I_B$, along the rung junctions. The gauge-invariant phase difference across rung junction $j$ is $\gamma_j$, while the phase difference across the off-rung junctions along the outer(inner) edge of plaquette $j$ is $\psi_{1,j}$($\psi_{2,j}$). The critical current, resistance, and capacitance of rung junction $j$ are denoted $I_{cj}$, $R_j$, and $C_j$, respectively. For simplicity, we assume all off-rung junctions are identical, with critical current $I_{co}$, resistance $R_o$, and capacitance $C_o$. We also assume that the product of the junction critical current and resistance is the same for all junctions in the array\cite{Benz95}, with a similar assumption about the ratio of each junction's critical current with its capacitance:
\begin{equation}
I_{cj}R_j=I_{co}R_o=\frac{\langle I_c\rangle}{\langle R^{-1}\rangle}
\label{eq:icR}
\end{equation}
\begin{equation}
\frac{I_{cj}}{C_j}=\frac{I_{co}}{C_o}=\frac{\langle I_c\rangle }{\langle C\rangle},
\label{eq:icC}
\end{equation}
where for any generic quantity $X$, the angular brackets with no subscript denote an arithmetic average over the set of rung junctions, $\langle X\rangle\equiv (1/N)\sum_{j=1}^{N}X_j$.

For convenience, we work with dimensionless quantities. Our dimensionless time variable is
\begin{equation}
\tau\equiv\frac{t}{t_c}=\frac{2e\langle I_c\rangle t}{\hbar\langle R^{-1}\rangle},
\label{eq:tau}
\end{equation}
where $t$ is the ordinary time. The dimensionless bias current is
\begin{equation}
i_B\equiv\frac{I_B}{\langle I_c\rangle},
\label{eq:iB}
\end{equation}
while the dimensionless critical current of rung junction $j$ is $i_{cj}\equiv I_{cj}/\langle I_c\rangle$. The McCumber parameter in this case is
\begin{equation}
\beta_c\equiv\frac{2e\langle I_c\rangle\langle C\rangle}{\hbar\langle R^{-1}\rangle^2}.
\label{eq:Mccumber}
\end{equation}
Note that $\beta_c$ is proportional to the mean capacitance of the rung junctions. An important dimensionless parameter is
\begin{equation}
\alpha\equiv\frac{I_{co}}{\langle I_c\rangle},
\label{eq:alpha}
\end{equation}
which will effectively tune the nearest-neighbor interaction strength in our phase model for the ladder.

Conservation of charge applied to the superconducting islands on the outer and inner edge, respectively, of rung junction $j$ yields the following equations in dimensionless variables:
\begin{subequations}
\begin{equation}
i_B - i_{cj}\sin\gamma_j - i_{cj}\frac{d\gamma_j}{d\tau} - i_{cj}\beta_c\frac{d^2\gamma_j}{d\tau^2} - \alpha\sin\psi_{1,j} - \alpha\frac{d\psi_{1,j}}{d\tau} - \alpha\beta_c\frac{d^2\psi_{1,j}}{d\tau^2}+ \alpha\sin\psi_{1,j-1} + \alpha\frac{d\psi_{1,j-1}}{d\tau} + \alpha\beta_c\frac{d^2\psi_{1,j-1}}{d\tau^2}=0,
\label{eq:outer}
\end{equation}
\begin{equation}
-i_B + i_{cj}\sin\gamma_j + i_{cj}\frac{d\gamma_j}{d\tau} + i_{cj}\beta_c\frac{d^2\gamma_j}{d\tau^2} - \alpha\sin\psi_{2,j} - \alpha\frac{d\psi_{2,j}}{d\tau} - \alpha\beta_c\frac{d^2\psi_{2,j}}{d\tau^2}+ \alpha\sin\psi_{2,j-1} + \alpha\frac{d\psi_{2,j-1}}{d\tau} + \alpha\beta_c\frac{d^2\psi_{2,j-1}}{d\tau^2}=0,
\label{eq:inner}
\end{equation}
\label{eq:conservation}
\end{subequations}
where $1\leq j\leq N$. The result is a set of $2N$ equations in $3N$ unknowns: $\gamma_j$, $\psi_{1,j}$, and $\psi_{2,j}$. We supplement Eq.~\ref{eq:conservation} by the constraint of fluxoid quantization in the absence of external or induced magnetic flux. For plaquette $j$ this constraint yields the relationship
\begin{equation}
\gamma_j + \psi_{2,j} - \gamma_{j+1} - \psi_{1,j} =0.
\label{eq:fluxoid}
\end{equation}
Equations~\ref{eq:conservation} and~\ref{eq:fluxoid} can be solved numerically for the $3N$ phases $\gamma_j$, $\psi_{1,j}$ and $\psi_{2,j}$\cite{comment1}.

We assign the rung junction critical currents in one of two ways, randomly or nonrandomly. We generate random critical currents according to a parabolic probability distribution function (pdf) of the form
\begin{equation}
P(i_c)=\frac{3}{4\Delta^3}\left[\Delta^2 - (i_c -1)^2\right],
\label{eq:randomic}
\end{equation}
where $i_c=I_c/\langle I_c\rangle$ represents a scaled critical current, and $\Delta$ determines the spread of the critical currents. Equation~\ref{eq:randomic} results in critical currents in the range $1-\Delta\leq i_c\leq 1+\Delta$. Note that this choice for the pdf (also used in Ref.~\cite{Wiesenfeld96}) avoids extreme critical currents (relative to a mean value of unity) that are occasionally generated by pdf's with tails. The nonrandom method of assigning rung junction critical currents was based on the expression
\begin{equation}
i_{cj}=1+\Delta -\frac{2\Delta}{(N-1)^2}\left[4j^2 - 4(N+1)j + (N+1)^2\right], \hspace{1em} 1\leq j\leq N,
\label{eq:quadraticic}
\end{equation}
which results in the $i_{cj}$ values varying quadratically as a function of position along the ladder and falling within the range $1-\Delta \leq i_{cj}\leq 1+\Delta$. We usually use $\Delta =0.05$.

\subsection{\label{sec:multipletime} Multiple time scale analysis}

Our goal in this subsection is to derive a Kuramoto-like model for the phase differences across the rung junctions, $\gamma_j$, starting with Eq.~\ref{eq:conservation}. We begin with two reasonable assumptions. First, we assume there is a simple phase relationship between the two off-rung junctions in the same plaquette:
\begin{equation}
\psi_{2,j}=-\psi_{1,j},
\label{eq:orphases}
\end{equation}
the validity of which has been discussed in detail elsewhere\cite{Daniels2003,Filatrella95}. As a result, Eq.~\ref{eq:fluxoid} reduces to
\begin{equation}
\psi_{1,j}=\frac{\gamma_j - \gamma_{j+1}}{2},
\label{eq:psi1}
\end{equation}
which implies that Eq.~\ref{eq:outer} can be written as
\begin{eqnarray}
i_{cj}\beta_c\frac{d^2\gamma_j}{d\tau^2} + i_{cj}\frac{d\gamma_j}{d\tau} + \frac{\alpha\beta_c}{2}\left[\frac{d^2\gamma_{j+1}}{d\tau^2}-2\frac{d^2\gamma_j}{d\tau^2} + \frac{d^2\gamma_{j-1}}{d\tau^2}\right] + \frac{\alpha}{2}\left[\frac{d\gamma_{j+1}}{d\tau} - 2\frac{d\gamma_j}{d\tau}+\frac{d\gamma_{j-1}}{d\tau}\right]= \nonumber \\
i_B - i_{cj}\sin\gamma_j + \alpha\sum_{\delta=\pm 1}\sin\left(\frac{\gamma_{j+\delta}-\gamma_j}{2}\right).
\label{eq:intermediate}
\end{eqnarray}

Our second assumption is that we can neglect the discrete Laplacian terms in Eq~\ref{eq:intermediate}, namely $\nabla^2(d\gamma_j/d\tau)\equiv d\gamma_{j+1}/d\tau - 2d\gamma_j/d\tau + d\gamma_{j-1}/d\tau$ and $\nabla^2(d^2\gamma_j/d\tau^2)\equiv d^2\gamma_{j+1}/d\tau^2 - 2d^2\gamma_j/d\tau^2 + d^2\gamma_{j-1}/d\tau^2$. We find numerically, over a wide range of bias currents $i_B$, McCumber parameters $\beta_c$, and coupling strengths $\alpha$ that $\nabla^2(d\gamma_j/d\tau)$ and $\nabla^2(d^2\gamma_j/d\tau^2)$ oscillate with a time-averaged value of approximately zero. Since the multiple time scale method is similar to averaging over a fast time scale, it seems reasonable to drop these terms. In light of this assumption, Eq.~\ref{eq:intermediate} becomes
\begin{equation}
i_{cj}\beta_c\frac{d^2\gamma_j}{d\tau^2} + i_{cj}\frac{d\gamma_j}{d\tau} = i_B - i_{cj}\sin\gamma_j + \alpha\sum_{\delta=\pm 1}\sin\left(\frac{\gamma_{j+\delta} - \gamma_j}{2}\right).
\label{eq:start}
\end{equation}

We can use Eq.~\ref{eq:start} as the starting point for a multiple time scale analysis. Following Refs.~\cite{Chernikov95} and~\cite{Watanabe97}, we divide Eq.~\ref{eq:start} by $i_B$ and define the following quantities:
\begin{subequations}
\label{eq:scalings}
\begin{equation}
\tilde{\tau}\equiv i_B\tau
\label{eq:newtau}
\end{equation}
\begin{equation}
\tilde{\beta_c}\equiv i_B\beta_c
\label{eq:newbetac}
\end{equation}
\begin{equation}
\epsilon=1/i_B.
\label{eq:epsilon}
\end{equation}
\end{subequations}
In terms of these scaled quantities, Eq.~\ref{eq:start} can be written as
\begin{equation}
1=i_{cj}\tilde{\beta_c}\frac{d^2\gamma_j}{d\tilde{\tau}^2} + i_{cj}\frac{d\gamma_j}{d\tilde{\tau}}+\epsilon i_{cj}\sin\gamma_j - \epsilon\alpha\sum_{\delta}\sin\left(\frac{\gamma_{j+\delta} - \gamma_j}{2}\right).
\label{eq:scaledstart}
\end{equation}
Next, we introduce a series of four (dimensionless) time scales,
\begin{equation}
T_n\equiv \epsilon^n \tilde{\tau} \hspace{1em} n=0,1,2,3,
\label{eq:timescales}
\end{equation}
which are assumed to be independent of each other. Note that $0<\epsilon < 1$ since $\epsilon = 1/i_B$. We can think of each successive time scale, $T_n$,  as being ``slower'' than the scale before it. For example, $T_2$ describes a slower time scale than $T_1$. The time derivatives in Eq.~\ref{eq:scaledstart} can be written in terms of the new time scales, since we can think of $\tilde{\tau}$ as being a function of the four independent $T_n$'s, $\tilde{\tau}=\tilde{\tau}(T_0,T_1,T_2,T_3)$. Letting $\partial_n\equiv\partial/\partial T_n$, the first and second time derivatives can be written as
\begin{equation}
\frac{d}{d\tilde{\tau}}=\partial_0 + \epsilon\partial_1 + \epsilon^2\partial_2 + \epsilon^3\partial_3
\label{eq:derivative1}
\end{equation}
\begin{equation}
\frac{d^2}{d\tilde{\tau}^2}=\partial^{2}_{0} + 2\epsilon\partial_0\partial_1 + \epsilon^2\left(2\partial_0\partial_2 + \partial^{2}_{1}\right) + 2\epsilon^3\left(\partial_0\partial_3 + \partial_1\partial_2\right),
\label{eq:derivative2}
\end{equation}
where in Eq.~\ref{eq:derivative2} we have dropped terms of order $\epsilon^4$ and higher. 

Next, we expand the phase differences in an $\epsilon$ expansion
\begin{equation}
\gamma_j = \sum_{n=0}^{\infty}\epsilon^n\gamma_{n,j}(T_0,T_1,T_2,T_3).
\label{eq:expansion}
\end{equation}
Substituting this expansion into Eq.~\ref{eq:scaledstart} and collecting all terms of order $\epsilon^0$ results in the expression
\begin{equation}
1=i_{cj}\tilde{\beta_c}\partial^{2}_{0}\gamma_{0,j} + i_{cj}\partial_{0}\gamma_{0,j},
\label{eq:order0}
\end{equation}
for which we find the solution
\begin{equation}
\gamma_{0,j}=\frac{T_0}{i_{cj}} + \phi_j(T_1,T_2,T_3),
\label{eq:gamma0}
\end{equation}
where we have ignored a transient term of the form $e^{-T_0/\tilde{\beta_c}}$, and where $\phi_j(T_1,T_2,T_3)$ is assumed constant over the fastest time scale $T_0$. Note that the expression for $\gamma_{0,j}$ consists of a rapid phase rotation described by $T_0/i_{cj}$ and slower-scale temporal variations, described by $\phi_j$, on top of that overturning. In essence, the goal of this technique is to solve for the dynamical behavior of the slow phase variable, $\phi_j$. The remaining details of the calculation can be found in the Appendix. We merely quote the resulting differential equation for the $\phi_j$ here:
\begin{eqnarray}
\beta_c\frac{d^2\phi_j}{d\tau^2}+\frac{d\phi_j}{d\tau}=\Omega_j + K_j\sum_{\delta=\pm 1}\sin\left[\frac{\phi_{j+\delta}-\phi_j}{2}\right] + L_j\sum_{\delta=\pm 1}\sin\left[3\left(\frac{\phi_{j+\delta}-\phi_j}{2}\right)\right] \nonumber \\
+M_j\sum_{\delta =\pm 1}\left\{\cos\left[\frac{\phi_{j+\delta}-\phi_j}{2}\right] -\cos\left[3\left(\frac{\phi_{j+\delta}-\phi_j}{2}\right)\right]\right\},
\label{eq:fullresult}
\end{eqnarray}
where $\Omega_j$ is given by the expression (letting $x_j\equiv i_{cj}/i_B$ for convenience)
\begin{equation}
\Omega_j=\frac{1}{x_j}\left[1 - \frac{x_{j}^{4}}{\left(2\beta_c^2+x_{j}^{2}\right)}\right],
\label{eq:omega}
\end{equation}
and the three coupling strengths are
\begin{equation}
K_j=\frac{\alpha}{i_{cj}}\left[1+\frac{x_{j}^{4}\left(3x_{j}^{2} + 23\beta_c^2\right)}{16\left(\beta_c^2 + x_{j}^{2}\right)^2}\right],
\label{eq:Kj}
\end{equation}
\begin{equation}
L_j=\frac{\alpha}{i_{cj}}\frac{x_{j}^{4}\left(3\beta_c^2-x_{j}^{2}\right)}{16\left(\beta_c^2+x_{j}^{2}\right)^2},
\label{eq:Lj}
\end{equation}
\begin{equation}
M_j=-\frac{\alpha}{i_{cj}}\frac{x_{j}^{5}\beta_c}{4\left(\beta_c^2+x_{j}^{2}\right)^2}.
\label{eq:Mj}
\end{equation}
We emphasize that Eq.~\ref{eq:fullresult} is expressed in terms of the original, unscaled, time variable $\tau$ and McCumber parameter $\beta_c$.

We will generally consider bias current and junction capacitance values such that $x_{j}^{2} \ll \beta_c^2$. In this limit, Eqs.~\ref{eq:Kj} -~\ref{eq:Mj} can be approximated as follows:
\begin{equation}
K_j\rightarrow \frac{\alpha}{i_{cj}}\left[ 1 + {\cal O}\left(\frac{1}{i_{B}^{4}}\right)\right],
\label{eq:approxKj}
\end{equation}
\begin{equation}
L_j\rightarrow \frac{\alpha}{i_{cj}}\left(\frac{3x_{j}^{4}}{16\beta_c^2}\right)\sim {\cal O}\left(\frac{1}{i_{B}^{4}}\right),
\label{eq:approxLj}
\end{equation}
\begin{equation}
M_j\rightarrow -\frac{\alpha}{i_{cj}}\left(\frac{x_{j}^{5}}{4\beta_c^3}\right)\sim {\cal O}\left(\frac{1}{i_{B}^{5}}\right).
\label{eq:approxMj}
\end{equation}
For large bias currents, it is reasonable to truncate Eq.~\ref{eq:fullresult} at ${\cal O}(1/i_{B}^{3})$, which leaves
\begin{equation}
\beta_c\frac{d^2\phi_j}{d\tau^2}+\frac{d\phi_j}{d\tau}=\Omega_j + \frac{\alpha}{i_{cj}}\sum_{\delta=\pm 1}\sin\left[\frac{\phi_{j+\delta}-\phi_j}{2}\right],
\label{eq:workingresult}
\end{equation}
where all the cosine coupling terms and the third harmonic sine term have been dropped as a result of the truncation. 

In the absence of any coupling between neighboring rung junctions ($\alpha =0$) the solution to Eq.~\ref{eq:workingresult} is
\[
\phi_{j}^{(\alpha =0)}= A + Be^{-\tau/\beta_c} + \Omega_j\tau,
\]
where $A$ and $B$ are arbitrary constants. Ignoring the transient exponential term, we see that $d\phi_{j}^{(\alpha =0)}/d\tau=\Omega_j$, so we can think of $\Omega_j$ as the voltage across rung junction $j$ in the uncoupled limit. Alternatively, $\Omega_j$ can be viewed as the angular velocity of the strongly-driven rotator in the uncoupled limit.

Equation~\ref{eq:workingresult} is our desired phase model for the rung junctions of the underdamped ladder. The result can be described as a locally-coupled Kuramoto model with a second-order time derivative (LKM2) and with junction coupling determined by $\alpha$. In the context of systems of coupled rotators, the second derivative term is due to the non-negligible rotator inertia, whereas in the case of Josephson junctions the second derivative arises because of the junction capacitance. The \textit{globally-coupled} version of the second-order Kuramoto model (GKM2) has been well studied; in this case the oscillator inertia leads to a first-order synchronization phase transition as well as to hysteresis between a weakly and a strongly coherent synchronized state\cite{Tanaka97,Acebron2000}.

\section{\label{sec:results}Comparison of LKM2 and RCSJ models}

We now compare the synchronization behavior of the RCSJ ladder array with the LKM2. We consider frequency and phase synchronization separately. For the rung junctions of the ladder, frequency synchronization occurs when the time average voltages, $\langle v_j\rangle_{\tau}=\langle d\phi_j/d\tau\rangle_{\tau}$ are equal for all $N$ junctions, within some specified precision. In the language of coupled rotators, this corresponds to phase points moving around the unit circle with the same average angular velocity. We quantify the degree of frequency synchronization via an ``order parameter''
\begin{equation}
f = 1-\frac{s_v(\alpha)}{s_v(0)},
\label{eq:f}
\end{equation}
where $s_v(\alpha)$ is the standard deviation of the $N$ time-average voltages, $\langle v_j\rangle_{\tau}$:
\begin{equation}
s_v(\alpha) = \sqrt{\frac{\sum_{j=1}^{N}\left(\langle v_j\rangle_{\tau} - \frac{1}{N}\sum_{k=1}^{N}\langle v_k\rangle_{\tau}\right)^2}{N-1}}
\label{eq:sv}
\end{equation}
In general, this standard deviation will be a function of the coupling strength $\alpha$, so $s_v(0)$ is a measure of the spread of the $\langle v_j\rangle_{\tau}$ values for $N$ independent junctions. Frequency synchronization of all $N$ junctions is signaled by $f=1$, while $f=0$ means all $N$ average voltages have their uncoupled values.

\begin{figure}
\includegraphics[scale=0.6]{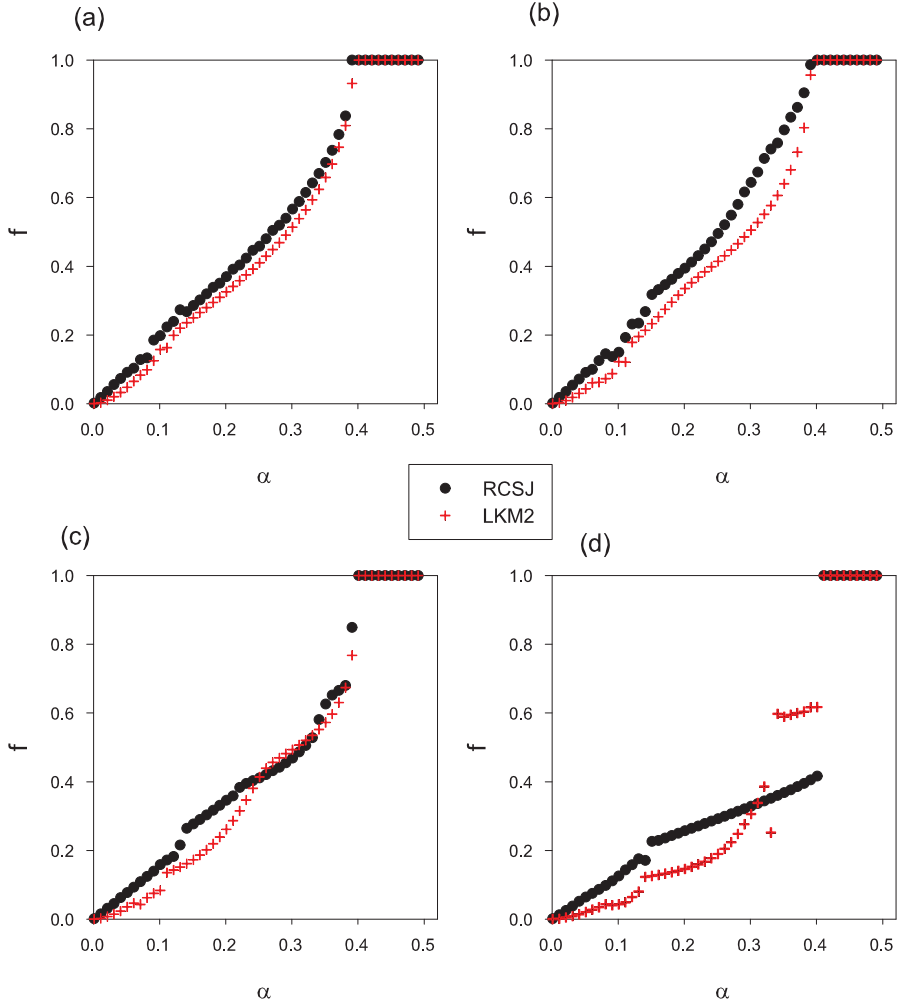}
\caption{\label{fig:fsync} (Color online) Frequency synchronization order parameter $f$, plotted versus nearest-neighbor coupling strength $\alpha$ for a ladder with $N=10$ plaquettes and bias current $i_B=5$. Rung junction critical currents are assigned nonrandomly with $\Delta =0.05$. (a) $\beta_c=1$, (b) $\beta_c=5$, (c) $\beta_c=10$, (d) $\beta_c=20$. For each plot the phase differences and voltages are reset to zero with each new value of $\alpha$.}
\end{figure}

Figure~\ref{fig:fsync} compares the order parameter $f$ for an array with $N=10$ plaquettes, a bias current of $i_B=5$, and nonrandomly-assigned critical currents with $\Delta =0.05$ for both the RCSJ model and the LKM2. For the RCSJ model, Eqs.~\ref{eq:conservation} and~\ref{eq:fluxoid} were solved numerically using a fourth-order Runge Kutta algorithm with a time step of $\Delta\tau=0.005$ and a total of $5\times 10^5$ time steps. All time-average quantities were evaluated using the second half of the time interval. For the LKM2, the same numerical approach was applied to Eq.~\ref{eq:workingresult}.
Figure~\ref{fig:fsync} shows some interesting behavior. First, in general, the LKM2 agrees well with the RCSJ model, especially in predicting a critical coupling strength, $\alpha_c$, at the onset of full frequency synchronization ($f=1$). Second, as $\beta_c$ is increased both models show evidence of a first order transition at $\alpha_c$ (see Fig.~\ref{fig:fsync}(d)) at which $f$ jumps abruptly to a value of unity. In the vicinity of such an abrupt transition, the models differ the most, but even in Fig.~\ref{fig:fsync}(d), the RCSJ model and the LKM2 agree on the value of $\alpha_c$. The deviation between the models seen in Fig.~\ref{fig:fsync}(d) near $\alpha\approx 0.4$ could be due to a region of bistability near $\alpha_c$ that becomes more prominent for increasing $\beta_c$.  

\begin{figure}
\includegraphics[scale=0.6]{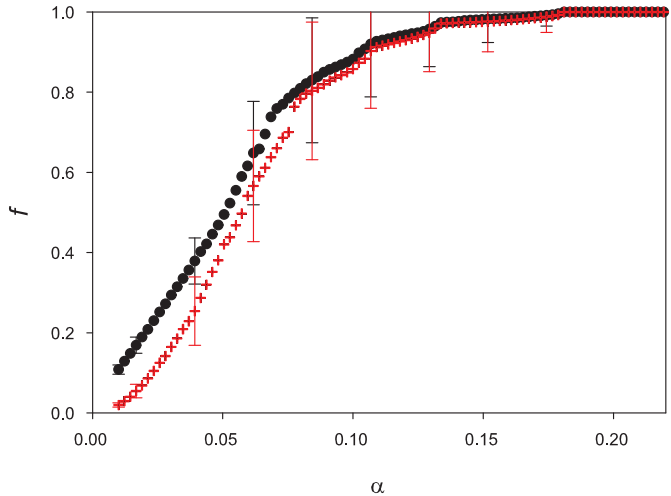}
\caption{\label{fig:fsyncrandom} (Color online) Frequency synchronization order parameter $f$, plotted versus nearest-neighbor coupling strength $\alpha$ for a ladder with $N=15$ plaquettes, bias current $i_B=5$, and $\beta_c =20$. Rung junction critical currents are assigned randomly with $\Delta =0.025$, and results are averaged over ten realizations of critical currents. Error bars represent the standard deviation of the mean value of $f$ and for clarity only a few, representative error bars are shown. Numerical solution of the RCSJ model are denoted by filled circles, and the results from the LKM2 are denoted by crosses.}
\end{figure}

Figure~\ref{fig:fsyncrandom} shows the case where the critical currents are assigned randomly according to Eq.~\ref{eq:randomic} with $\Delta=0.025$ for $N=15$, $i_B=5$, and $\beta_c=20$. The results for the frequency synchronization order parameter were obtained by averaging over ten different critical current realizations, and the error bars are the standard deviation of the mean value of $f$ for each $\alpha$. Note the excellent agreement between the RCSJ model and the LKM2. Also note that averaging over critical current realizations has a smoothing effect of $f$ compared to, for example, Fig.~\ref{fig:fsync}(d).

\begin{figure}
\includegraphics[scale=0.6]{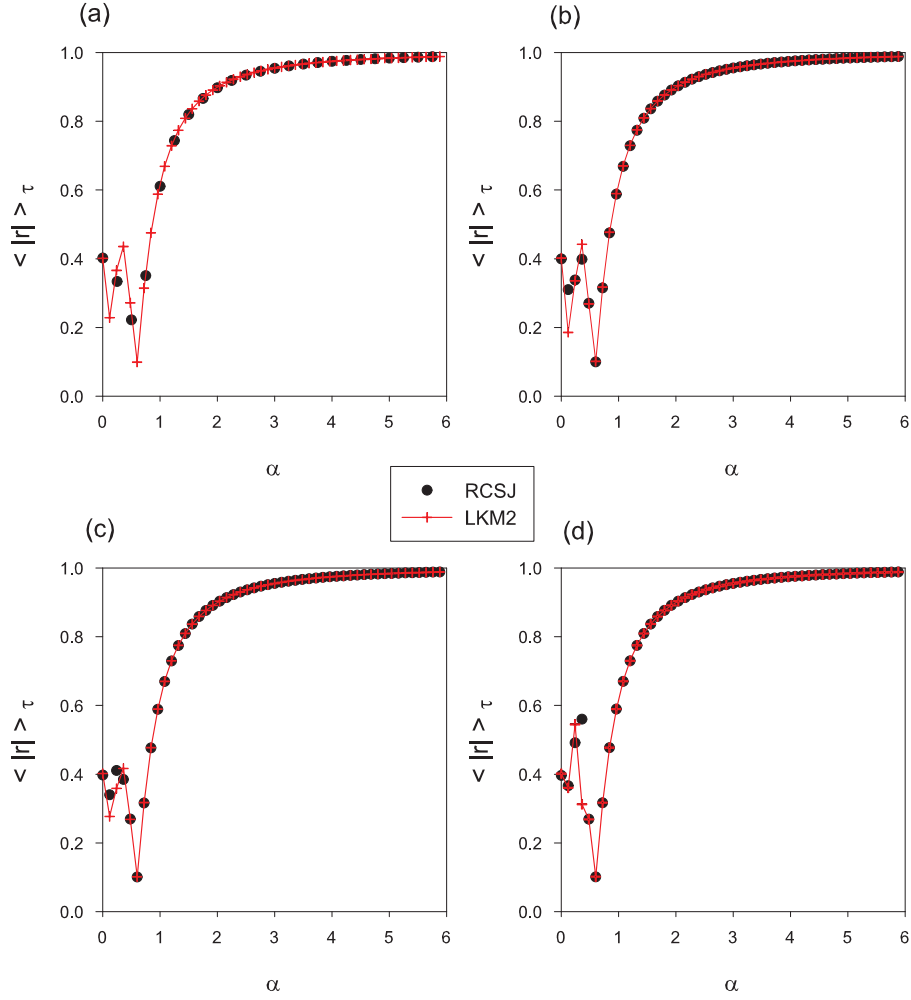}
\caption{\label{fig:rsync} (Color online) Phase synchronization order parameter $\langle |r|\rangle_{\tau}$, plotted versus nearest-neighbor coupling strength $\alpha$ for a ladder with $N=10$ plaquettes, $i_B=5$, and non-random critical currents with $\Delta =0.05$. (a) $\beta_c=1$, (b) $\beta_c=5$, (c) $\beta_c=10$, (d) $\beta_c=20$. For each plot phase differences and voltages across rung junctions are reset to zero with each new value of $\alpha$.}
\end{figure}

Phase synchronization of the rung junctions is measured by the usual Kuramoto order parameter
\begin{equation}
r\equiv \frac{1}{N}\sum_{j=1}^{N} e^{i\phi_j}.
\label{eq:r}
\end{equation}
The results shown in Fig.~\ref{fig:rsync} represent the time-averaged modulus of $r$, $\langle |r|\rangle_{\tau}$, which approaches unity when the phase differences across the junctions are identical. Figure~\ref{fig:rsync} compares the phase synchronization of the RCSJ model and the LKM2 for the same geometry as in Fig.~\ref{fig:fsync}. The agreement between the two models is excellent. Note the two types of behavior observable in the plots. For small coupling ($\alpha \alt 0.7$), $\langle |r|\rangle_{\tau}$ displays a complicated behavior due to finite-size effects, while for $\alpha \agt 0.7$, $\langle |r|\rangle_{\tau}$ exhibits a smooth rise toward a value of unity with increasing coupling. In fact, comparison of Figs.~\ref{fig:fsync} and~\ref{fig:rsync} shows that the value of $\alpha$ signaling the onset of the smooth increase in phase synchronization is approximately equal to $\alpha_c$, the value at which full frequency synchronization is obtained. Figure.~\ref{fig:rsync}(d) also suggests that the finite-size fluctuations for small $\alpha$ are more pronounced at large $\beta_c$ (compare with Figs.~\ref{fig:rsync}(a), (b), (c)). Since the second-order Kuramoto model with \textit{global} coupling (GKM2) has discontinuities in $\langle |r|\rangle_{\tau}$ as a function of coupling strength for large arrays\cite{Tanaka97}, and since we have mapped the RCSJ model to the LKM2, it would be interesting to look for evidence of a first-order transition in $\langle |r|\rangle_{\tau}$ for large arrays. Such evidence is already visible, even for arrays as small as $N=10$, in the frequency synchronization order parameter (see Fig.~\ref{fig:fsync}(d)).

\begin{figure}
\includegraphics[scale=0.6]{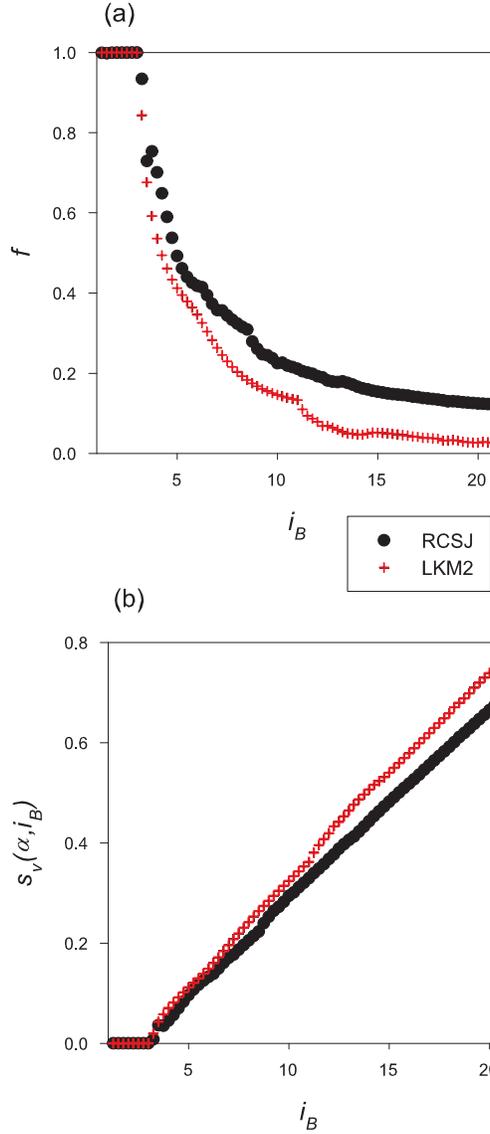}
\caption{\label{fig:ibsync} (Color online) (a) Frequency synchronization order parameter $f$, plotted versus dc bias current $i_B$ for fixed coupling strength $\alpha=0.25$ for both the RCSJ model and the LKM2. $N=10$, $\beta_c=5$, and nonrandom critical currents with $\Delta =0.05$. (b) Standard deviation of time-averaged rung junction voltages versus bias current for the same model parameters as in (a). The difference between the two models evident at large bias currents is probably due to phase slips in the off-rung junctions that violate Eq.~\ref{eq:orphases}.}
\end{figure}

We have also studied the synchronization in our two models as a function of the dc bias current $i_B$ for fixed coupling $\alpha$, as shown in Fig.~\ref{fig:ibsync}. Such a graph is useful because experiments on periodic ladders would most likely be performed at fixed $\alpha$ (since that quantity is set by the fabrication of the rung and off-rung junctions), while the bias current could be easily varied. To obtain $f$ experimentally, then, one needs to measure the time-average voltages across the rung junctions for each value of the bias current. Figure~\ref{fig:ibsync}(a) demonstrates that as the bias current is increased for fixed coupling strength, frequency synchronization is eventually lost. This is reasonable physically; as a rotator is driven harder a stronger coupling with its neighbors should be required to keep the rotators entrained. Figure~\ref{fig:ibsync}(b) plots $s_v(\alpha,i_B)$ versus $i_B$, showing that the spread in junction voltages scales linearly with the bias current over a wide range of currents. The behavior observed in both Figs.~\ref{fig:ibsync}(a) and (b) for bias currents of $i_B\agt 10$ is not surprising. When the system is far from frequency synchronization, the time-averaged voltages should be well approximated by their values in the absence of coupling, namely $\langle v_j\rangle_{\tau}\approx \Omega_j$, where $\Omega_j$ is given by Eq.~\ref{eq:omega}. In the limit $(i_{cj}/i_B)^2\ll 1$, Eq.~\ref{eq:omega} gives $\Omega_j\approx i_B/i_{cj}$. In this case, we can write
\begin{equation}
s_v(\alpha,i_B)\approx i_B\sqrt{\frac{\sum_{j=1}^{N}\left(\frac{1}{i_{cj}}-\frac{1}{N}\sum_{k=1}^{N}\frac{1}{i_{ck}}\right)^2}{N-1}} = Ci_B,
\label{eq:svlimit}
\end{equation}
where $C$ is a constant independent of the bias current. Thus the linear scaling of $s_v$ with bias current is just due to the scaling of the time-averaged voltages across the rung junctions with $i_B$. Equation~\ref{eq:svlimit} is actually the standard deviation in the limit $\alpha\rightarrow 0$, so for bias currents large enough that the junctions can be treated as approximately independent, we expect $s_v(\alpha,i_B)/s_v(0,i_B)\rightarrow 1$, which in turn means $f\rightarrow 0$, as observed in Fig.~\ref{fig:ibsync}(a).

\begin{figure}
\includegraphics[scale=0.75]{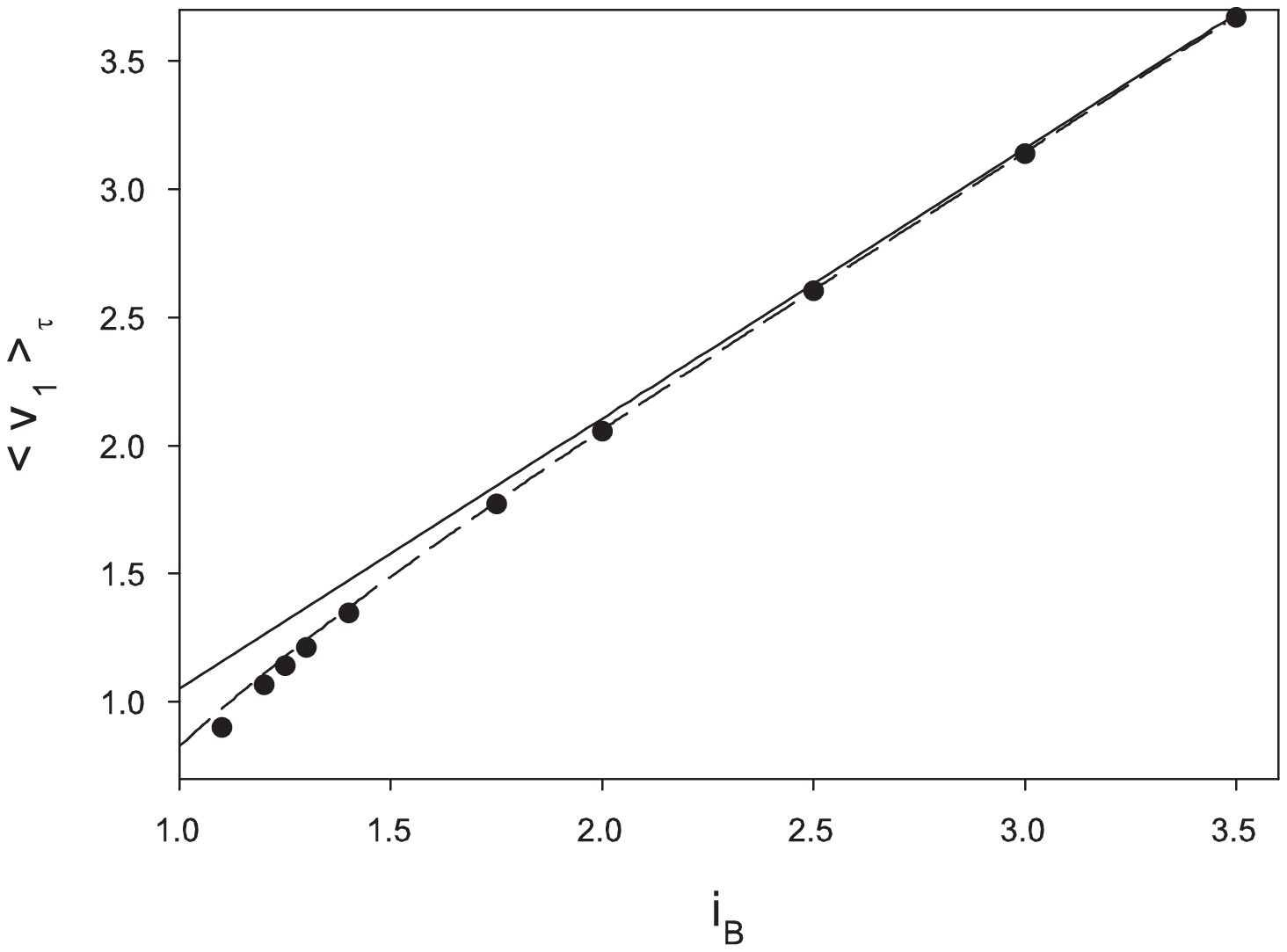}
\caption{\label{fig:<v1>} Time-averaged voltage across $j=1$ rung junction, plotted versus bias current $i_B$ for $N=10$, $\alpha = 0.01\ll\alpha_c$, and $\beta_c =1$. The solid circles are a numerical result from the RCSJ model. They agree well with the analytical result for the weak-coupling limit, $\langle v_1\rangle_{\tau}=\Omega_1$, where $\Omega_1$ is given by Eq.~\ref{eq:omega}, which is represented by a dotted line in the graph. Deviations from the large bias current result, $\langle v_1\rangle_{\tau}\approx i_B/i_{c1}$ (solid line), do not appear until $i_B\alt 2.5.$}
\end{figure}

Figure~\ref{fig:<v1>} shows that $\lim_{\alpha\rightarrow 0}\langle v_j\rangle_{\tau}=\Omega_j$. To obtain this result, $\langle v_j\rangle_{\tau}(i_B)$ across the $j=1$ rung junction was calculated numerically for the RCSJ model for $\beta_c=1$ and $\alpha=0.01$, which is more than an order of magnitude smaller than $\alpha_c$. The results are shown as solid circles. The dotted line represents the analytic expression $\langle v_1\rangle_{\tau} = \Omega_1$, where $\Omega_1$ is given by Eq.~\ref{eq:omega} and which results from our multiple time-scale analysis. The solid line is the large bias current limit of Eq.~\ref{eq:omega}, namely $\Omega_1\approx i_B/i_{c1}$. Note that the numerical results agree well with the Eq.~\ref{eq:omega} for $\alpha\ll\alpha_c$ over the entire range of bias currents shown, and with the large bias current result for $i_B\agt 2.5$.

\begin{figure}
\includegraphics[scale=0.6]{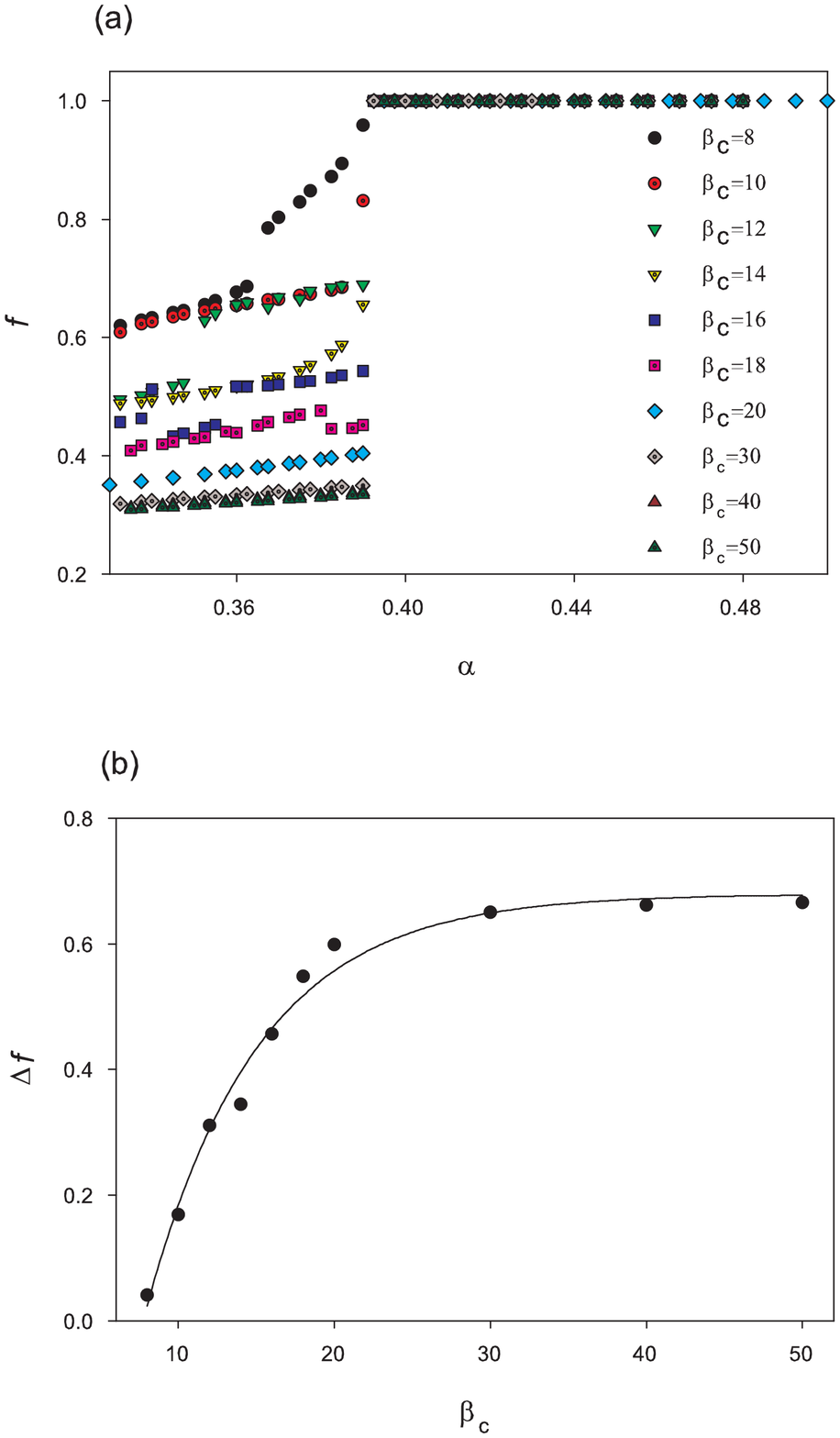}
\caption{\label{fig:gap} (a) (Color online) Frequency synchronization order parameter $f$, plotted versus coupling strength $\alpha$ for a ladder with $N=10$, $i_B=5$, and nonrandom critical currents with $\Delta =0.05$. Results based on the RCSJ model for ten different $\beta_c$ values are shown. In these simulations, $\alpha$ is initialized to a value greater than $\alpha_c$ and then gradually decreased. With each new value of $\alpha$ the phase differences and voltages across the rung junctions are reset to zero. Note the appearance of a discontinuity in $f$ as $\beta_c$ is increased. There is also some evidence of additional discontinuities in $f$ for $\alpha <\alpha_c$ (see, for example, the data for $\beta_c =8, 12$). (b) Discontinuity in the frequency synchronization order parameter $\Delta f$ at $\alpha=\alpha_c$ for different values of $\beta_c$. This graph is produced from the data for $f$ versus $\alpha$ seen in (a). The discontinuity is well fit by an exponential rise to a maximum, $\Delta f=\Delta f_{max}\left[1-\exp{-b\left(\beta_c -\beta_{c}^{*}\right)}\right]$, where we find parameter values of $\Delta f_{max}=0.679 \pm 0.020$, $b=0.140\pm0.020$ and $\beta_{c}^{*}=7.750\pm 0.294$ for this array.}
\end{figure}

\begin{figure}
\includegraphics[scale=0.6]{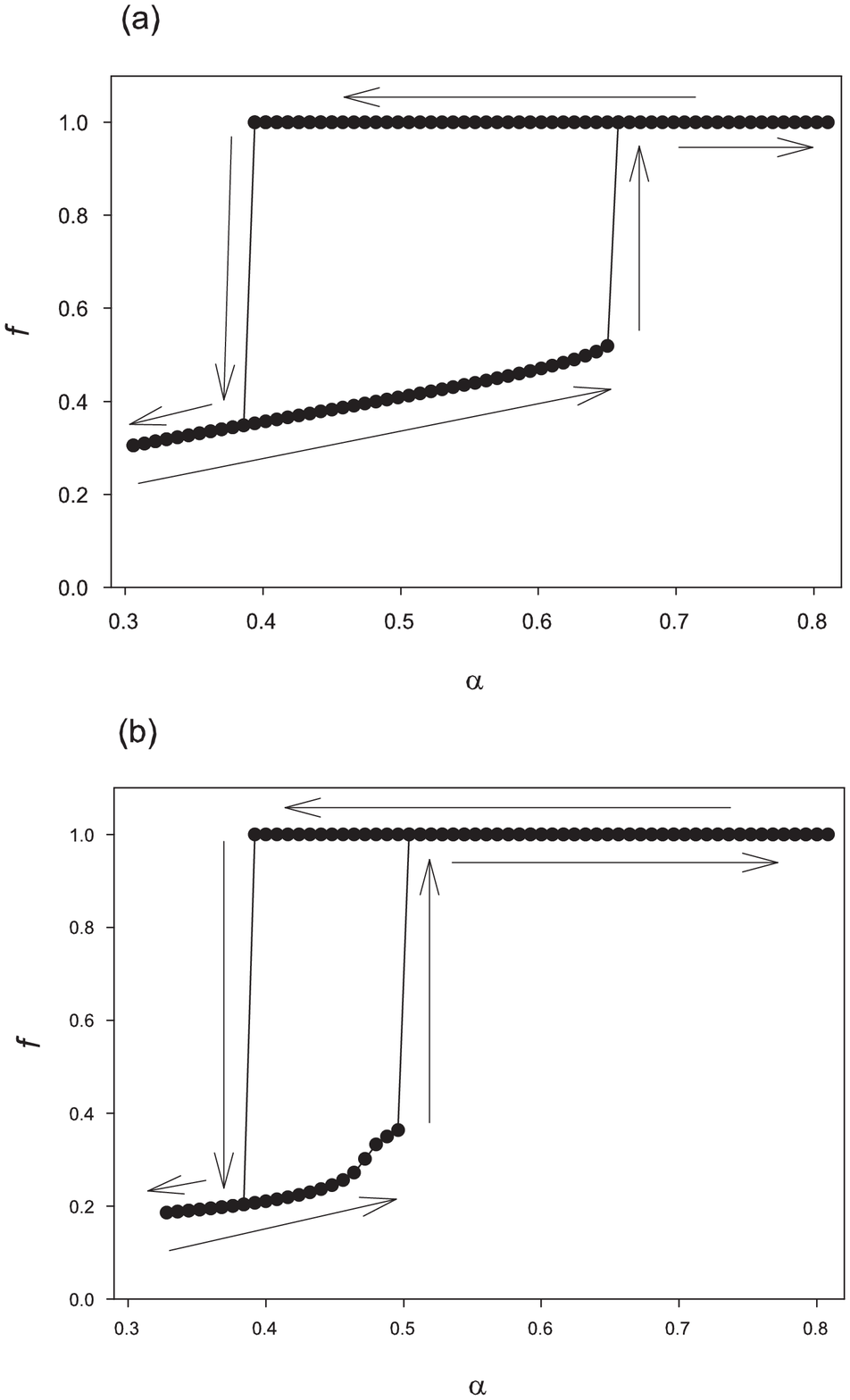}
\caption{\label{fig:hysteresis} Hysteresis in the frequency synchronization order parameter $f$, plotted as a function of the coupling strength $\alpha$ for a ladder with $N=10$, $i_B=5$, $\beta_c =30$, and nonrandom critical currents with $\Delta =0.05$. To produce these plots, $\alpha$ is initialized to a value greater than $\alpha_c$ and then gradually decreased until the system jumps discontinuously to the unsynchronized state; the coupling $\alpha$ is then increased until the system jumps back to the synchronized state. The final values of the phase differences and voltages across the junctions, as obtained from the previous value of $\alpha$, are used as the initial values for each new value of $\alpha$. (a) Results based on the RCSJ model. (b) Results based on the LKM2. Note that the models agree on the value of $\alpha_c$ at which the jump occurs from the synchronized to the unsynchronized state for decreasing $\alpha$. The solid lines are a guide to the eye.}
\end{figure}

\begin{figure}
\includegraphics[scale=0.6]{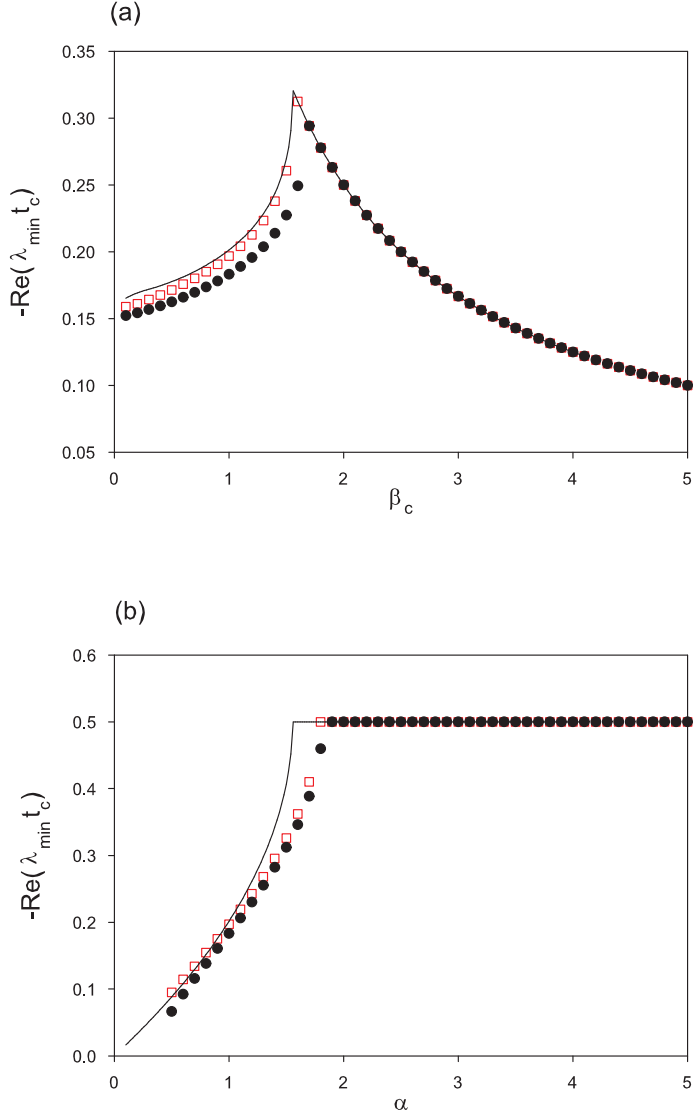}
\caption{\label{fig:floquet} (Color online) Real part of the minimum magnitude Floquet exponent for an array with $N=10$ and $i_B=5$. (a) Dependence of exponents on $\beta_c$ for fixed coupling strength, $\alpha =1$. Symbols are results of a numerical calculation based on the RCSJ model with either no disorder (open squares) or non-random critical currents and $\Delta=0.05$ (solid circles). The solid line is an analytic result (Eq.~\ref{eq:fexponents}) from the LKM2. (b) Dependence of exponents on coupling strength $\alpha$ for fixed $\beta_c=1$.}
\end{figure}

Of particular interest is how the array behaves near the frequency synchronization transition, $\alpha\approx\alpha_c$ . As shown in Fig.~\ref{fig:gap}(a) for an array with $N=10$ plaquettes driven by a bias current $i_B=5$, the order parameter develops a discontinuity at $\alpha_c$ for $\beta_c \agt 8$. In addition, for sufficiently large $\beta_c$ the array also exhibits hysteretic behavior in $f$, as shown in Fig.~\ref{fig:hysteresis}. The behavior depicted in Figs.~\ref{fig:gap} and~\ref{fig:hysteresis} is presumably due to bistability of the individual junctions\cite{Tanaka97b} arising from their non-negligible capacitance. Figure~\ref{fig:gap}(b) shows that for increasing $\beta_c$ the discontinuity in the order parameter at $\alpha_c$, $\Delta f$, can be well fit by an exponential rise that asymptotically saturates to a value $\Delta f_{max}$ for $\beta_c\rightarrow\infty$, \textit{i.e.}
\begin{equation}
\Delta f = \left\{\begin{array}{ll}
      \Delta f_{max}\left[1 - e^{-b(\beta_c - \beta_{c}^{*})}\right] & \beta_c \geq \beta_{c}^{*} \\
      0 & \beta_c < \beta_{c}^{*}.
     \end{array}
    \right.
\label{eq:gap}
\end{equation}
For $\beta_c < \beta_{c}^{*}$ the frequency synchronization transition is a smooth function of $\alpha$ as $\alpha$ decreases through $\alpha_c$ from above. Thus, the junctions must be sufficiently underdamped for the discontinuous nature of the transition to be manifest. (But Fig.~\ref{fig:gap}(a) gives at least the hint of a possible discontinuity in $\Delta f$ for $\beta_c=8$ around $\alpha=0.37 < \alpha_c$.) The data in Fig.~\ref{fig:gap} were obtained from a numerical solution of the RCSJ model, but we see qualitatively similar behavior from a numerical solution of Eq.~\ref{eq:workingresult}, namely saturation of $\Delta f$ to a maximum value that is well fit by an exponential function.
Lastly in this section, we address the issue of the linear stability of the frequency synchronized states ($\alpha > \alpha_c$) by calculating their Floquet exponents numerically for the RCSJ model as well as analytically based on the LKM2, Eq.~\ref{eq:workingresult}. The analytic technique used has been described in detail elsewhere\cite{Trees2001}, so we shall merely quote the result for the real part of the Floquet exponents:
\begin{equation}
\mbox{Re}(\lambda_m t_c)=-\frac{1}{2\beta_c}\left[1\pm\mbox{Re}\sqrt{1-4\beta_c \left(\bar{K}+3\bar{L}\right)\omega_{m}^{2}}\right],
\label{eq:fexponents}
\end{equation}
where stable solutions correspond to exponents, $\lambda_m$, with a negative real part. One can think of the $\omega_m$ as the normal mode frequencies of the ladder. We find that for a ladder with periodic boundary conditions and $N$ plaquettes
\begin{equation}
\omega_{m}^{2}=\frac{4\sin^2\left(\frac{m\pi}{N}\right)}{1+2\sin^2\left(\frac{m\pi}{N}\right)}, \hspace{2em} 0 \leq m\leq N-1.
\label{eq:normalmodes}
\end{equation}
To arrive at Eq.~\ref{eq:fexponents} we have ignored the effects of disorder so that $\bar{K}$ and $\bar{L}$ are obtained from Eqs.~\ref{eq:Kj} and~\ref{eq:Lj} with the substitution $i_{cj}\rightarrow 1$ throughout. This should be reasonable for the levels of disorder we have considered ($5\%$). Substituting the expressions for $\bar{K}$ and $\bar{L}$ into Eq.~\ref{eq:fexponents} results in 
\begin{equation}
\mbox{Re}(\lambda_m t_c)=-\frac{1}{2\beta_c}\left[1\pm\mbox{Re}\sqrt{1-2\beta_c\alpha\left\{1 + \frac{2\beta_{c}^{2}}{\left(i_{B}^{2}\beta_{c}^{2}+1\right)^2}\right\}\omega_{m}^{2}}\right].
\label{eq:fexponents2}
\end{equation}
We are most interested in the Floquet exponent of minimum magnitude, $\mbox{Re}(\lambda_{\mbox{min}}t_c)$, which essentially gives the lifetime of the longest-lived perturbations to the synchronized state (see Fig.~\ref{fig:floquet}).

If the quantity inside the square root in Eq.~\ref{eq:fexponents2} is negative then $\mbox{Re}(\lambda_mt_c)=-1/(2\beta_c)$. This is the value seen in Fig.~\ref{fig:floquet}(a) for $\beta_c > \bar{\beta_c}=1.56$. For $\beta_c < \bar{\beta_c}$ and $m=1$, the quantity inside the square root is positive and $\mbox{Re}(\lambda_{\mbox{min}}t_c)=(-1/2\beta_c)\left[1-\sqrt{1-2\beta_c\alpha\omega_{1}^{2}}\right]$, where we have used the fact that the quantity inside the braces in Eq.~\ref{eq:fexponents2} is essentially unity for $i_B=5$ and $\beta_c > 1$. Physically, the crossover-type behavior evident in Fig.~\ref{fig:floquet}(a) is due to the low frequency (long wavelength), $m=1$, normal mode of the ladder changing from underdamped to overdamped in character as $\beta_c$ is decreased through $\bar{\beta_c}=1/(2\alpha\omega_{1}^{2})=1.56$ for $N=10$ and $\alpha =1$. Note from Fig.~\ref{fig:floquet}(a) that the numerical result for the exponents based on the RCSJ model, with $5\%$ disorder, agree quite well with Eq.~\ref{eq:fexponents2}. Not surprisingly, the RCSJ model with no critical current disorder agrees very well with the analytic result since the disorder was ignored in order to obtain Eq.~\ref{eq:fexponents2}. Figure~\ref{fig:floquet}(b) shows how the minimum-magnitude Floquet exponent varies with coupling strength $\alpha$ for fixed $\beta_c$. Now there is a crossover at $\alpha=\bar{\alpha}=1/(2\beta_c\omega_{1}^{2})=1.56$ for $\beta_c=1$.  For $\alpha > \bar{\alpha}$, Eq.~\ref{eq:fexponents2} gives $\mbox{Re}(\lambda_{\mbox{min}}t_c)=-1/(2\beta_c)$, independent of $\alpha$. For $\alpha < \bar{\alpha}$, however, we see that as $\alpha\rightarrow\alpha_c$ from above, the stability of the synchronized state decreases. In fact, one can show from Eq.~\ref{eq:fexponents2} that for $\alpha > \alpha_c$ and $\beta_c\alpha\omega_{1}^{2}\ll 1$, the stability decreases linearly with $\alpha$ according to
\[
\mbox{Re}\left(\lambda_{\mbox{min}}t_c\right)\approx -\frac{\alpha\omega_{1}^{2}}{2}, \hspace{1em}\alpha\rightarrow\alpha_c^+
\]
independent of $\beta_c$. Such linear behavior is evident in Fig.~\ref{fig:floquet}(b) for small $\alpha$.

\section{\label{sec:SW} ``Small-world'' connections in ladder arrays}

Many properties of small world networks have been studied in the last several years, including not only the effects of network topology but also the dynamics of the node elements comprising the network\cite{Newman2000,Strogatz2001}. Of particular interest has been the ability of oscillators to synchronize when configured in a small-world manner. Such synchronization studies can be broadly sorted into several categories.  (1) Work on coupled lattice maps has demonstrated that synchronization is made easier by the presence of random, long-range connections\cite{Gade2000,Batista2003}. (2) Much attention has been given to the synchronization of continuous time dynamical systems, including the first order locally-coupled Kuramoto model (LKM), in the presence of small world connections\cite{Hong2002a,Hong2002b,Watts99}. For example, Hong and coworkers\cite{Hong2002a,Hong2002b} have shown that the LKM, which does not exhibit a true dynamical phase transition in the thermodynamic limit ($N\rightarrow\infty$) in the \textit{pristine} case, does exhibit such a phase synchronization transition for even a small number of shortcuts. But the assertion\cite{Wang2002} that any small world network can synchronize for a given coupling strength and large enough number of nodes, even when the pristine network would not synchronize under the same conditions, is not fully accepted\cite{comment2}. (3) More general studies of synchronization in small world and scale-free networks\cite{Barahona2002,Nishikawa2003} have shown that the small world topology does not guarantee that a network can synchronize. In Ref.~\cite{Barahona2002} it was shown that one could calculate the average number of shortcuts per node, $s_{sync}$, required for a given dynamical system to synchronize. This study found no clear relation between this synchronization threshold and the onset of the small world region, \textit{i.e.} the value of $s$ such that the average path length between all pairs of nodes in the array is less than some threshold value. Reference~\cite{Nishikawa2003} studied arrays with a power-law distribution of node connectivities (scale-free networks) and found that a broader distribution of connectivities makes a network \textit{less} synchronizable even though the average path length is smaller. It was argued that this behavior was caused by an increased number of connections on the hubs of the scale-free network. Clearly it is dangerous to assume that merely reducing the average path length between nodes of an array will make such an array easier to synchronize.

How do Josephson junction arrays fit into the above discussion? Specifically, if we have a disordered array biased such that some subset of the junctions are in the voltage state, \textit{i.e.} undergoing limit cycle oscillations, will the addition of random, long-range connections between junctions aid the array in attaining frequency and/or phase synchronization? Our goal in this section of the paper is to address this question by using the mapping discussed in Secs.~\ref{sec:twotiming} and~\ref{sec:results} between the RCSJ model for the \textit{underdamped} ladder array and the second-order, locally-coupled Kuramoto model (LKM2). Based on the results of Ref.~\cite{Daniels2003}, we also know that the RSJ model for an \textit{overdamped} ladder can be mapped onto  a first-order, locally-coupled Kuramoto model (LKM). Because of this mapping, the ladder array falls into category (2) of the previous paragraph. In other words, we should expect the existence of shortcuts to drastically improve the ability of ladder arrays to synchronize.

\begin{figure}
\includegraphics[scale=0.60]{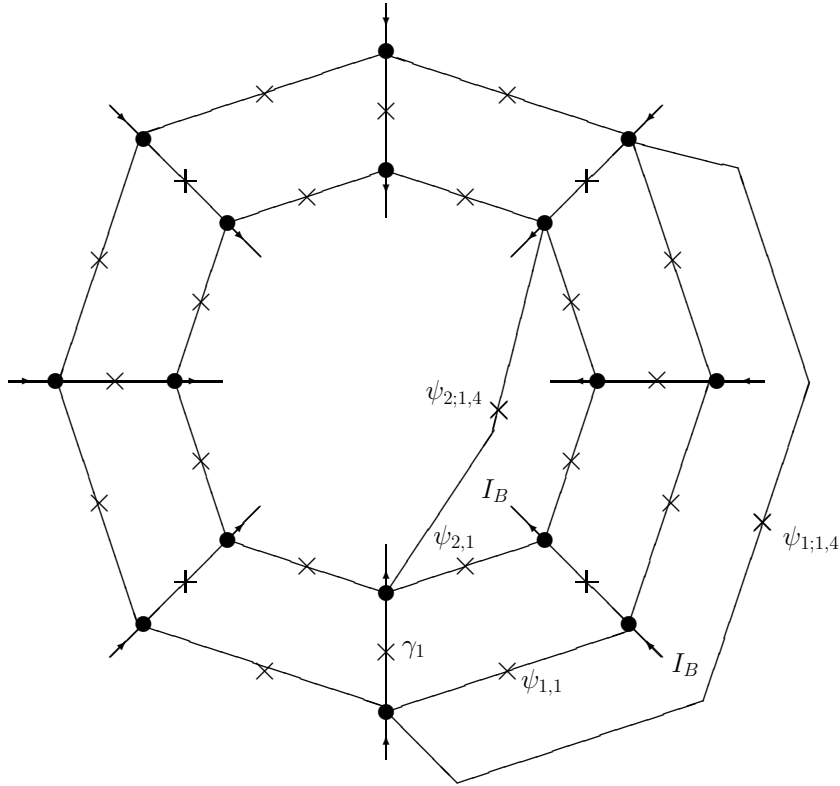}
\caption{\label{fig:SCladder} A ladder array with periodic boundary conditions, $N=8$ plaquettes, and one long-range connection. Rung junctions $j=1$ and $j=4$ are shown as connected by a pair of off-rung junctions, which have phase differences of $\psi_{1:1,4}$ and $\psi_{2;1,4}$. The additional off-rung junctions are uniform and identical to the off-rung junctions in the pristine ladder. Uniform dc bias currents are applied along the rungs as in Fig.~\ref{fig:ladder}.}
\end{figure}

We add connections between pairs of rung junctions that will result in interactions that are longer than nearest neighbor in range. We do so by adding two, nondisordered, off-rung junctions for each such connection. For example, Fig.~\ref{fig:SCladder} shows a connection between rung junctions $j=1$ and $j=4$. This is generated by the addition of the two off-rung junctions labeled $\psi_{1;1,4}$ and $\psi_{2:1,4}$, where the last two indices in each set of subscripts denote the two rung junctions connected. The new off-rung junctions are assumed to be identical to the original off-rung junctions in the array, with critical current $I_{co}$, resistance $R_o$, and capacitance $C_o$ for the underdamped case. Physically, we should expect that the new connection will provide a sinusoidal phase coupling between rung junctions $j=1$ and $j=4$ with a strength tuned by the parameter $\alpha=I_{co}/\langle I_c\rangle$, where $\langle I_c\rangle$ is the arithmetic average of the rung junction critical currents.  We assign long range connections between pairs of rung junctions randomly with a probability $p$ distributed uniformly between zero and one, and we do not allow multiple connections between the same pair of junctions. For the pristine ladder, with $p=0$, each rung has only nearest-neighbor connections, while $p=1$ corresponds to a regular network of globally coupled rung junctions, \textit{i.e.} each rung junction is coupled to every other rung junction in the ladder.

\begin{figure}
\includegraphics[scale=0.75]{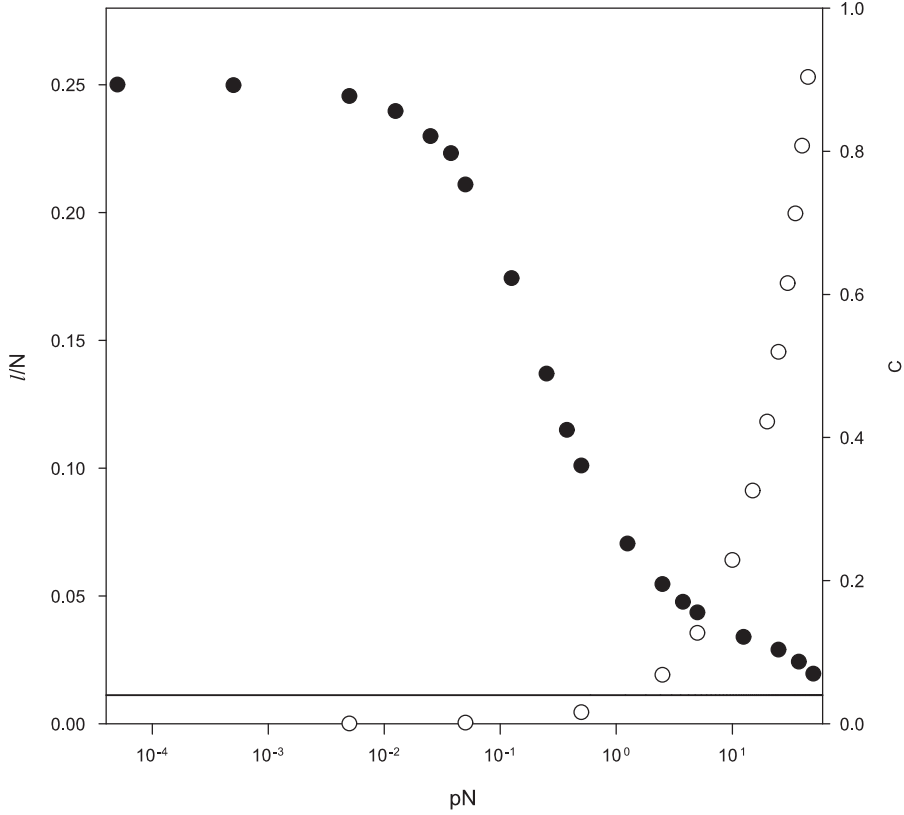}
\caption{\label{fig:l&c} Scaled average path length $l/N$ (solid circles and left vertical axis) and cluster coefficient $C$ (hollow circles and right vertical axis), plotted versus the product $pN$, where $p$ is the probability of making a shortcut connection between pairs of nodes and $N$ is the total number of nodes in the network. This graph corresponds to $N=50$. The solid line is $C=2/N$, which is the expected cluster coefficient for a random network of size $N=50$ in which each node has two neighbors in the pristine limit. Note that $C\rightarrow 0$ as $pN\rightarrow 0$, since each node's neighbors are not connected to each other in that limit. Also note that $C\rightarrow 1$ as $p\rightarrow 1$, which is the limit of a fully-connected, regular network in which each node's neighbors are all connected to one another.}
\end{figure}

In Fig.~\ref{fig:l&c} we plot the two standard quantities used to characterize the topology of the network: the average path length $l$ and the cluster coefficient $C$, calculated numerically for a network with $N=50$ nodes (\textit{i.e.} rung junctions). The average path length $l$ is defined as the minimum distance between each pair of nodes averaged over all such pairs, while the cluster coefficient $C$ is the average fraction of nodes neighboring each node that are also neighbors themselves. These quantities are plotted as a function of the product $pN$.  For $pN\approx 1$, the average path length is already substantially reduced from its value in the pristine limit, $p=0$. It is this reduced average distance between pairs of nodes that is one of the hallmarks of a small-world network. Because our ladder array, in the pristine limit, allows only nearest-neighbor coupling, the cluster coefficient vanishes as $p\rightarrow 0$. As a result, our ladder geometry does not conform to the most commonly-accepted definition of a small world, in which both reduced path lengths (compared to the pristine limit) and high cluster coefficients (compared to that of a random network) coexist (see Fig.~2 in Ref.~\cite{Watts98}). Nevertheless, in our simulations we routinely choose value for the parameter $p$ such that $pN$ places us in the region of reduced path lengths. So we are considering ladder arrays in which the average distance between rung junctions is reduced by shortcuts by a factor of five to ten.

\begin{figure}
\includegraphics[scale=0.6]{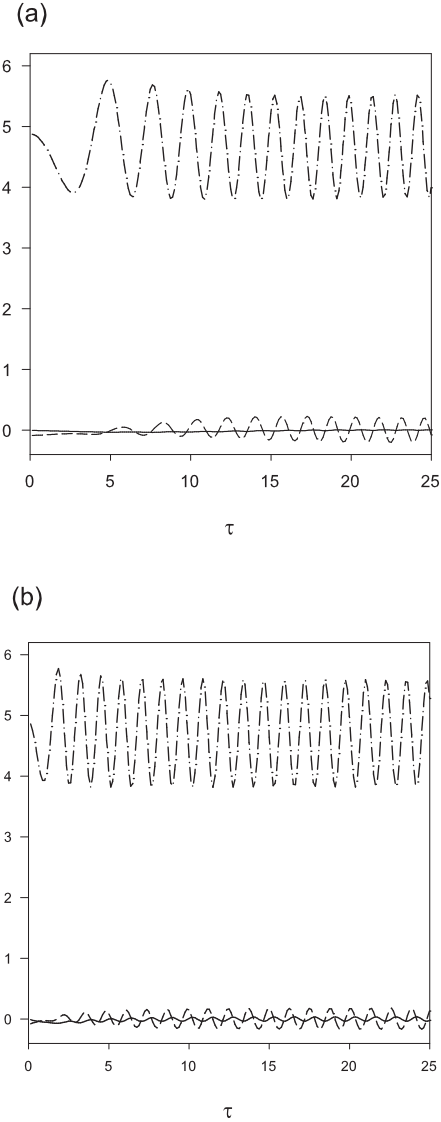}
\caption{\label{fig:approxcheck} Time dependence of several combinations of voltages or voltage derivatives for a ladder with $N=10$, $i_B=5$, and nonrandom critical currents with $\Delta =0.05$. The array was chosen to have three shortcuts between the following pairs of rung junctions: (1,3), (2,7), (4,6). In both plots the following quantities are compared (note that $v_1=d\gamma_1/d\tau$): $i_{c1}v_1 + i_{c1}\beta_c dv_1/d\tau$ (dot-dashed line), $(\alpha/2)(v_1 - v_3)$ (solid line), and $(\alpha/2)\beta_c(dv_1/d\tau - dv_3/d\tau)$ (dashed line). The time average value of the latter two quantities is negligible. (a) $\alpha=1$, $\beta_c=10$. (b) $\alpha=1$, $\beta_c=1$. }
\end{figure}

Next, we argue that the RCSJ equations for the underdamped junctions in the ladder array can be mapped onto a straightforward variation of 
Eq.~\ref{eq:workingresult}, in which the sinusoidal coupling term for rung junction $j$ also includes the longer-range couplings due to the added shortcuts. Imagine a ladder with a shortcut between junctions $j$ and $l$, where $l\neq j,j\pm 1$. Conservation of charge applied to the two superconducting islands that comprise rung junction $j$ will lead to equations very similar to Eq.~\ref{eq:conservation}. For example, the analog to Eq.~\ref{eq:outer} will be
\begin{eqnarray}
i_B - i_{cj}\sin\gamma_j - i_{cj}\frac{d\gamma_j}{d\tau} - \beta_ci_{cj}\frac{d^2\gamma_j}{d\tau^2} - \alpha\sin\psi_{1,j} - \alpha\frac{d\psi_{1,j}}{d\tau} -\beta_c\alpha\frac{d^2\psi_{1,j}}{d\tau^2}+ \\ \nonumber
  \alpha\sin\psi_{1,j-1}+\alpha\frac{d\psi_{1,j-1}}{d\tau}+\beta_c\alpha\frac{d^2\psi_{1,j-1}}{d\tau^2} +\sum_{l}\left[\alpha\sin\psi_{1;jl}+\alpha\frac{d\psi_{1;jl}}{d\tau}+\beta_c\alpha\frac{d^2\psi_{1;jl}}{d\tau^2}\right]=0, 
\label{eq:outerSC}
\end{eqnarray}
with an analogous equation corresponding to the inner superconducting island that can be generalized from Eq.~\ref{eq:inner}. The sum over the index $l$ accounts for all junctions connected to junction $j$ via an added shortcut. Fluxoid quantization still holds, which means that we can augment Eq.~\ref{eq:fluxoid} with
\begin{equation}
\gamma_j +\psi_{2;jl} -\gamma_l -\psi_{1;jl}=0.
\label{eq:newfluxoid}
\end{equation}
We also assume the analog of Eq.~\ref{eq:orphases} holds:
\begin{equation}
\psi_{2;jl}=-\psi_{1;jl}.
\label{eq:neworphases}
\end{equation}
Equations~\ref{eq:newfluxoid} and~\ref{eq:neworphases} allow us to write the analog to Eq.~\ref{eq:psi1} for the case of shortcut junctions:
\begin{equation}
\psi_{1;jl}=\frac{\gamma_j - \gamma_l}{2}
\label{eq:newpsi1}
\end{equation}
Equation~\ref{eq:outerSC}, in light of Eq.~\ref{eq:newpsi1}, can be written as
\begin{eqnarray}
i_B -i_{cj}\sin\gamma_j -i_{cj}\frac{d\gamma_j}{d\tau} -\beta_ci_{cj}\frac{d^2\gamma_j}{d\tau^2} + \alpha\sum_{\delta=\pm 1}\sin\left(\frac{\gamma_{j+\delta} - \gamma_j}{2}\right) +\alpha\sum_{l}\sin\left(\frac{\gamma_j -\gamma_l}{2}\right) + \\ \nonumber
\frac{\alpha}{2}\nabla^2\left(\frac{d\gamma_j}{d\tau}\right) + \frac{\alpha}{2}\nabla^2\left(\frac{d^2\gamma_j}{d\tau^2}\right) +\frac{\alpha}{2}\sum_{l}\left(\frac{d\gamma_j}{d\tau}-\frac{d\gamma_l}{d\tau}\right) +\frac{\alpha}{2}\sum_{l}\left(\frac{d^2\gamma_j}{d\tau^2}-\frac{d^2\gamma_l}{d\tau^2}\right)=0,
\label{eq:newoutersc}
\end{eqnarray}
where the sums $\Sigma_{l}$ are over all rung junctions connected to $j$ via an added shortcut. As we did with the pristine ladder, we will drop the two discrete Laplacians, since they have a very small time average compared to the terms $i_{cj}d\gamma_j/d\tau +i_{cj}\beta_c d^2\gamma_j/d\tau^2$. The same is also true, however, of the terms $\alpha/2\sum_{l}(d\gamma_j/d\tau -d\gamma_l/d\tau)$ and $\alpha/2\sum_{l}(d^2\gamma_j/d\tau^2 - d^2\gamma_l/d\tau^2)$, as direct numerical solution of the full RCSJ equations in the presence of shortcuts demonstrates (see Fig.~\ref{fig:approxcheck}). So we shall drop these terms as well. Then Eq.~\ref{eq:newoutersc} becomes
\begin{equation}
i_B -i_{cj}\sin\gamma_j -i_{cj}\frac{d\gamma_j}{d\tau} -\beta_ci_{cj}\frac{d^2\gamma_j}{d\tau^2} + \frac{\alpha}{2}\sum_{k\in\Lambda_j}\sin\left(\frac{\gamma_k -\gamma_j}{2}\right),
\label{eq:reduced}
\end{equation}
where the sum is over all junctions in $\Lambda_j$, which is the set of all junctions connected to junction $j$. Based on our work in Sec.~\ref{sec:twotiming}, we can predict that a multiple time scale analysis of Eq.~\ref{eq:reduced} results in a phase model of the form
\begin{equation}
\beta_c\frac{d^2\phi_j}{d\tau^2} + \frac{d\phi_j}{d\tau} = \Omega_j + \frac{\alpha}{2}\sum_{k\in\Lambda_j}\sin\left(\frac{\phi_k -\phi_j}{2}\right),
\label{eq:LKMSW2}
\end{equation}
where $\Omega_j$ is give by Eq.~\ref{eq:omega}.
A similar analysis for the \textit{overdamped} ladder leads to the result
\begin{equation}
\frac{d\phi_j}{d\tau} = \Omega_{j}^{(1)} + \frac{\alpha}{2}\sum_{k\in\Lambda_j}\sin\left(\frac{\phi_k -\phi_j}{2}\right),
\label{eq:LKMSW}
\end{equation}
where the time-averaged voltage across each overdamped rung junction in the uncoupled limit is
\begin{equation}
\Omega_{j}^{(1)}=\sqrt{\left(\frac{i_B}{i_{cj}}\right)^2 -1}.
\label{eq:omegaoverdamped}
\end{equation}

\begin{figure}
\includegraphics[scale=0.6]{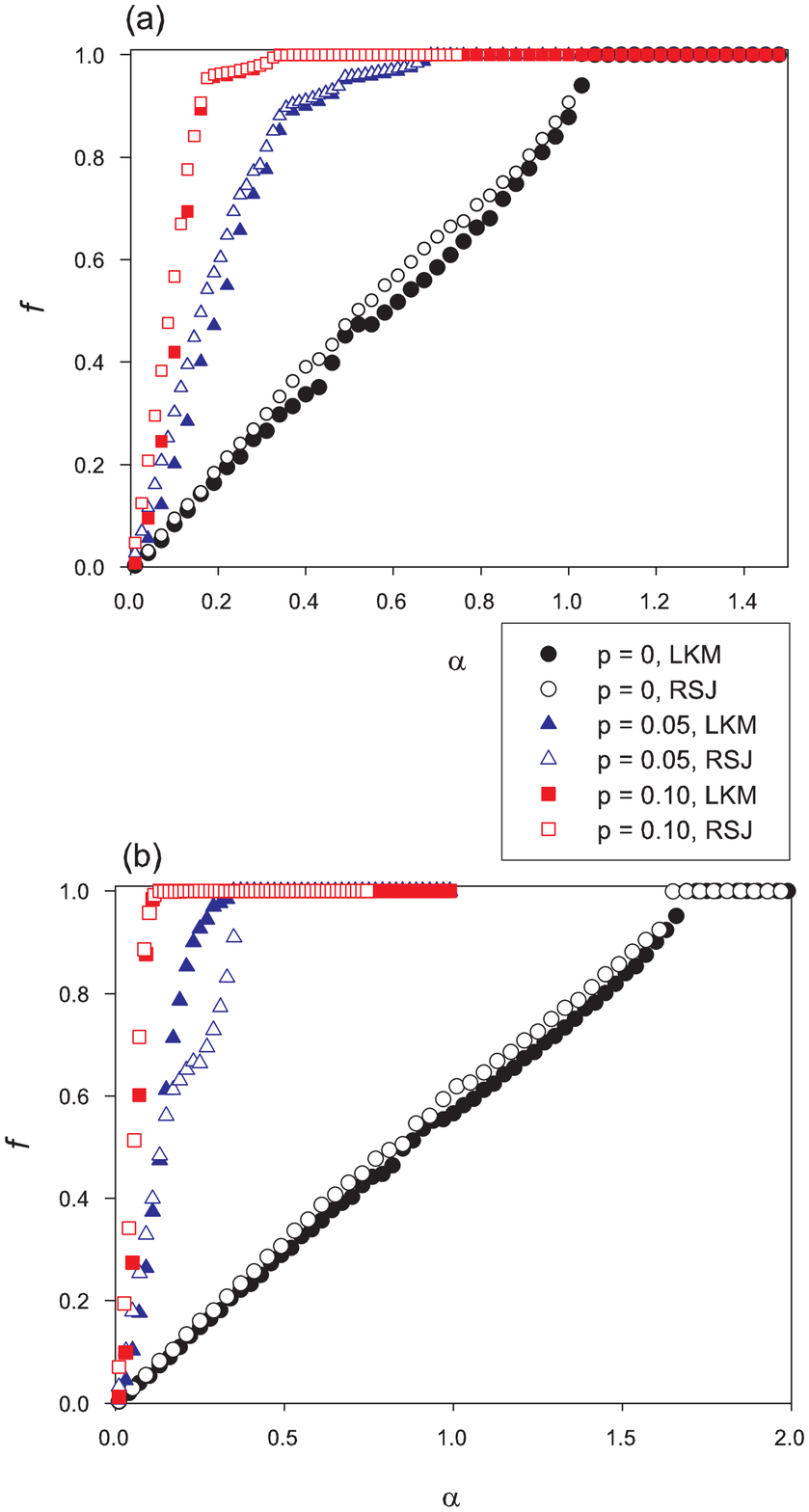}
\caption{\label{fig:3050overdamped} (Color online) Frequency synchronization order parameter $f$, plotted versus coupling strength $\alpha$ for \textit{overdamped} arrays with bias current $i_B=5$ and non-random critical currents with $\Delta =0.05$. Solid symbols represent numerical results based on the LKM model, Eq.~\ref{eq:LKMSW}, while hollow symbols are from the RSJ model. The data for nonzero $p$ represent an average over ten realizations of randomly-assigned shortcuts. (a) $N=30$, (b) $N=50$. Note that shortcuts improve the frequency synchronization behavior in that the critical coupling needed for $f=1$ is clearly reduced as $p$ is increased from zero. In fact $\alpha_c$ is clearly growing with increasing $N$ in the case of $p=0$, while the same is not true of the array with shortcuts.}
\end{figure}

\begin{figure}
\includegraphics[scale=0.6]{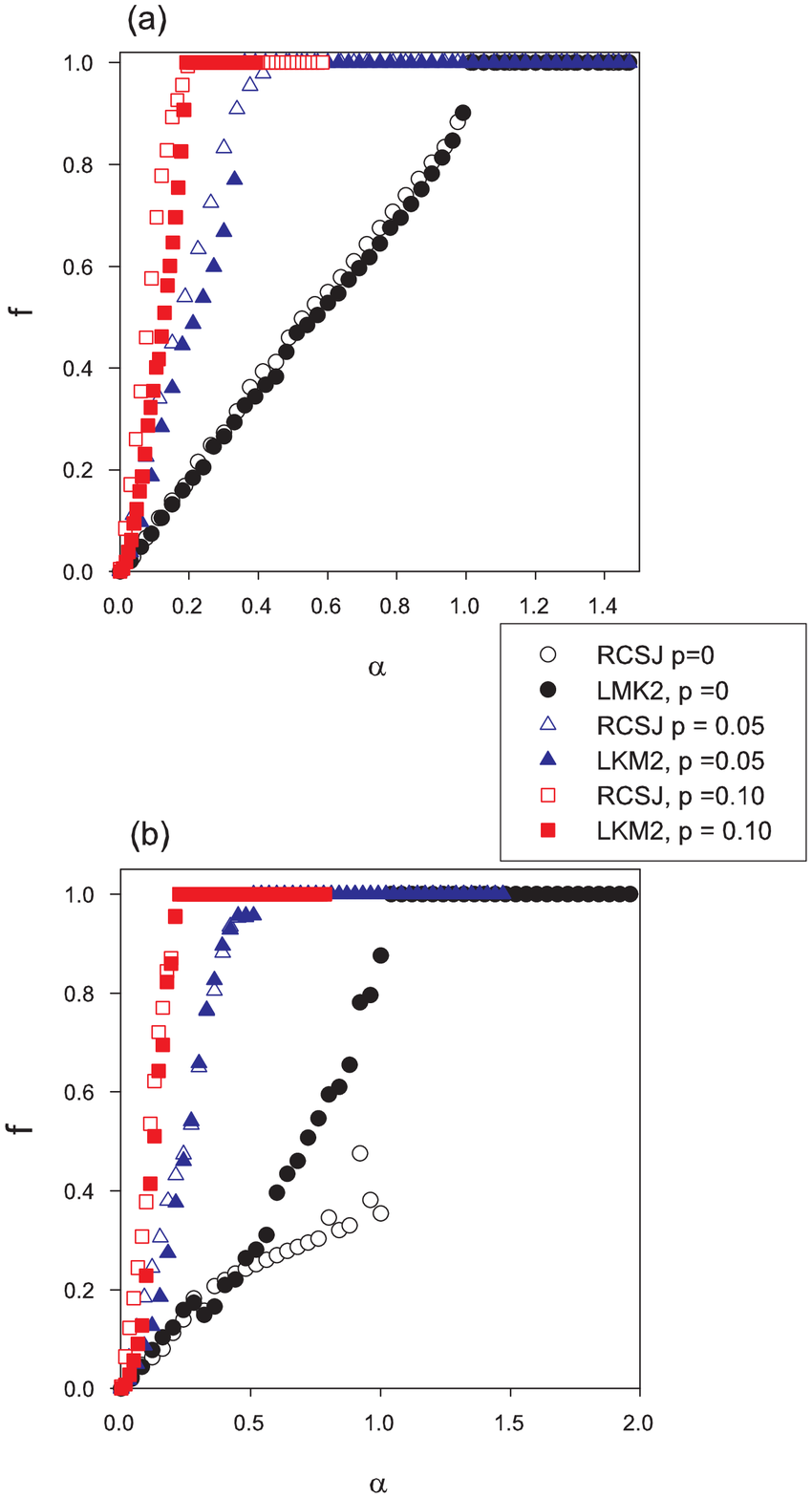}
\caption{\label{fig:3050underdamped} (Color online) Frequency synchronization order parameter $f$, plotted versus coupling strength $\alpha$ for an \textit{underdamped} ladder array with $N=30$, bias current $i_B=5$ and non-random critical currents with $\Delta=0.05$. Solid symbols represent numerical results based on the LKM2 model, Eq.~\ref{eq:LKMSW2}, while hollow symbols are from the RCSJ model. The data for nonzero $p$ represent an average over ten realizations of randomly-assigned shortcuts. (a) $\beta_c=1$, (b) $\beta_c=30$. Agreement between the two models is excellent, except for large $\beta_c$ and $p=0$ near $\alpha_c$, which is not unexpected based on Fig.~\ref{fig:fsync}.}
\end{figure}

Figure~\ref{fig:3050overdamped} demonstrates that the frequency synchronization order parameter $f$, calculated from Eq.~\ref{eq:LKMSW} for overdamped arrays with $N=30$ and $N=50$ and in the presence of shortcuts, agrees well with the results of the RSJ model. In addition to the pristine array with $p=0$, we considered arrays with $p=0.05$ and $p=0.10$ in which we averaged over 10 realizations of shortcuts. The agreement between the two models is excellent, as seen in the figure. It is also clear from the figure that shortcuts do indeed help frequency synchronization in that a smaller coupling strength $\alpha$ is required to reach $f=1$ in the presence of shortcuts. In fact, the value of $\alpha_c$ required to reach $f=1$ is growing with increasing $N$ in the cases of $p=0$; for example, we find the $\alpha_c=3.36$ for $N=100$ (compare with Fig.~\ref{fig:3050overdamped}). The same is clearly not true, however, for arrays with $p=0.05$ and $p=0.1$. Recently, Hong, Choi, and Kim\cite{Hong2002a} have demonstrated, using a finite-size scaling analysis applied to the LKM, that the phase synchronization order parameter, $\langle |r|\rangle_{\tau}$, in the presence of shortcuts ($0 < p < 1$) has a mean-field synchronization phase transition as in the GKM (\textit{i.e.} LKM with $p=1$). Based on the agreement between the two models shown in Fig.~\ref{fig:3050overdamped}, we expect such an analysis to apply to the RSJ equations for the ladder as well. The finite-size scaling behavior of the \textit{frequency} synchronization of the LKM has not been studied, so the nature of that transition is not well known. Figure~\ref{fig:3050underdamped} demonstrates that \textit{underdamped} ladders also synchronize more easily with shortcuts and that Eq.~\ref{eq:LKMSW2} agrees well with the RCSJ model.

\begin{figure}
\includegraphics[scale=0.75]{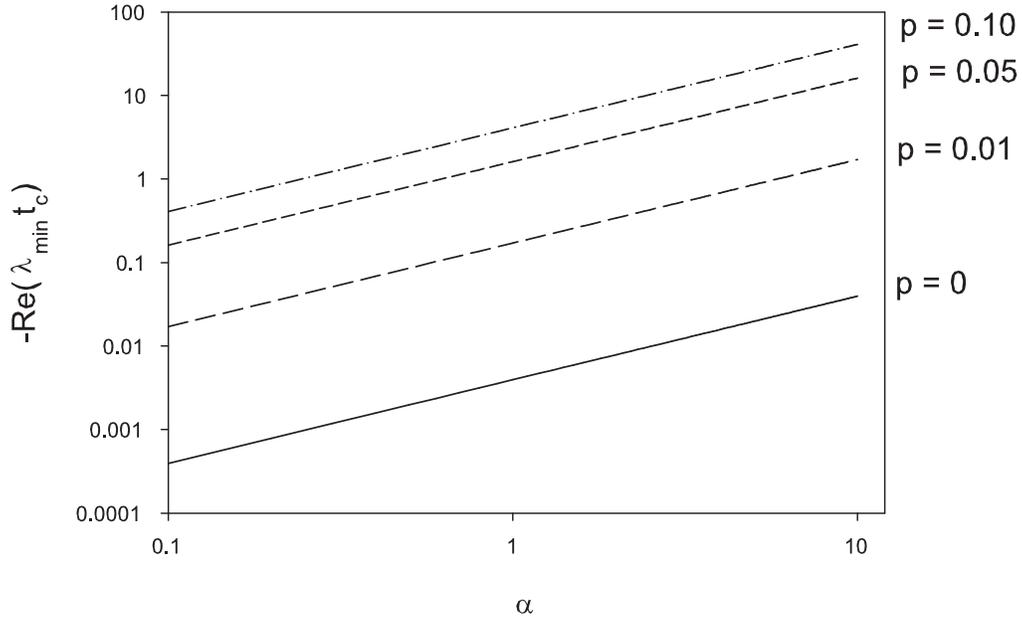}
\caption{\label{fig:floquetSC} Floquet exponent of minimum magnitude $\lambda_{\mbox{min}}$, plotted versus coupling strength $\alpha$ for the LKM with $N=100$. Exponents are calculated based on a technique described in Ref.~\cite{Pecora98}. For $p=0$, the exponents can be calculated analytically, with the result $\mbox{Re}(\lambda_{\mbox{min}}t_c)=-4\alpha\sin^2\left(\pi/N\right)$. For each nonzero $p$, the results are averaged over 100 realizations of shortcuts. Note the logarithmic scale on both axes.}
\end{figure}

Although the addition of shortcuts makes it easier for the array to synchronize, we should also consider the effects of such random connections on the stability of the synchronized state. The Floquet exponents for the synchronized state allow us to quantify this stability. Using a general technique discussed in Ref.~\cite{Pecora98}, we can calculate the Floquet exponents $\lambda_m$ for the LKM based on the expression
\begin{equation}
\lambda_mt_c = \alpha E_{m}^{G},
\label{eq:lambdaSC}
\end{equation}
where $E_{m}^{G}$ are the eigenvalues of \textbf{G}, the matrix of coupling coefficients for the array. A specific element, $G_{ij}$, of this matrix is unity if there is a connection between rung junctions $i$ and $j$. The diagonal terms, $G_{ii}$, is merely the negative of the number of junctions connected to junction $i$.  This gives the matrix the property $\sum_{j}G_{ij}=0$. In the case of the pristine ladder, the eigenvalues of \textbf{G} can be calculated analytically, which yields Floquet exponents of the form
\begin{equation}
\lambda{_m}^{(p=0)}t_c=-4\alpha\sin^2\left(\frac{m\pi}{N}\right).
\label{eq:lambdaSCp0}
\end{equation}
This result is plotted in Fig.~\ref{fig:floquetSC} as the solid line for an overdamped array with $N=100$; note that the solid line is the $p=0$ Floquet exponent of minimum, nonzero magnitude. Since the $E^{G}_{m}$ are purely geometry dependent, \textit{i.e.} do not depend on the coupling strength, we expect the exponents to grow linearly with $\alpha$, based on Eq.~\ref{eq:lambdaSCp0}. To include the effects of shortcuts, we found the eigenvalues $E^{G}_{m}$ numerically for a particular realization of shortcuts (for a given value of $p$), and then we averaged over 100 realizations of shortcuts for each value of $p$. The exponents of minimum magnitude for the overdamped array with $p=0.01, 0.05, 0.10$ are also shown in Fig.~\ref{fig:floquetSC} (note the logarithmic scale on both axes). Clearly, shortcuts greatly improve the stability of the synchronized state. Specifically $\lambda^{(p=0.1)}_{\mbox{min}}/\lambda^{(p=0)}_{\mbox{min}}=1030,$ a three-order of magnitude enhancement.

\section{\label{sec:2D} ``Small-world'' connections in two-dimensional arrays}

\begin{figure}
\includegraphics[scale=0.50]{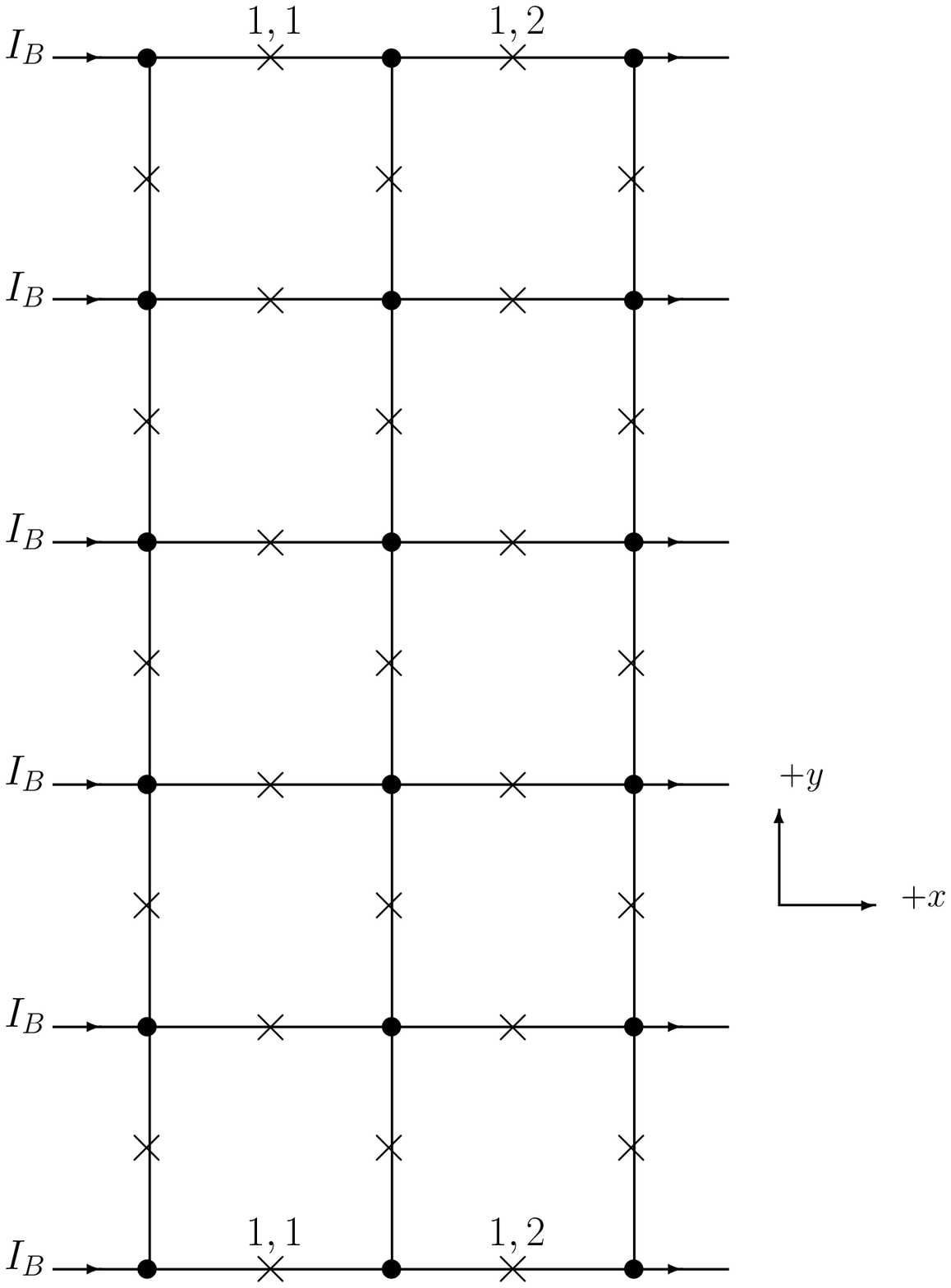}
\caption{\label{fig:2darray} A two dimensional array of junctions with $M=2$ columns, or ``ladders'', each with $N=5$ plaquettes. A dc bias current $I_B$ is injected along the left side and extracted along the ride side of the array. We assume periodic boundary conditions in the $y$ direction. A junction with a label such as $1,2$ means the junction is in row $1$ and column $2$.}
\end{figure}

\begin{figure}[t]
\includegraphics[scale=0.6]{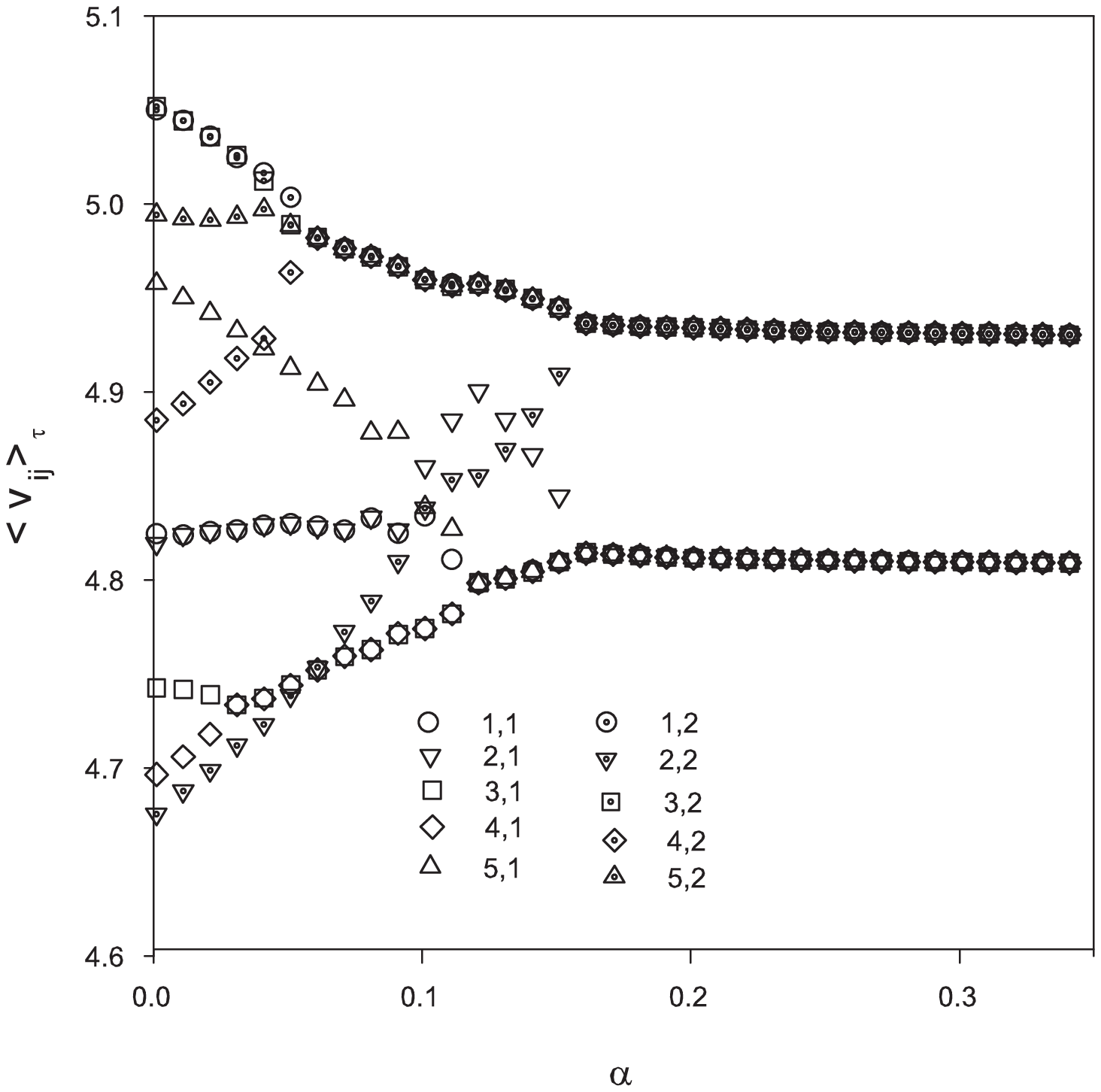}
\caption{\label{fig:cluster} A cluster diagram of the time-averaged voltages, $\langle v_{ij}\rangle_{\tau}$, across the horizontal junctions, plotted versus coupling strength $\alpha$ for an array with $M=2$ columns and $N=5$ plaquettes per column. A bias current $i_B=5$ is applied and extracted along each row, and the horizontal junctions are assigned critical currents randomly according to Eq.~\ref{eq:randomic} with $\Delta =0.05$. We assume periodic boundary conditions in the vertical direction. The array is pristine ($p=0$).  The symbols with dots correspond to voltages across the junctions in column 2. The diagram demonstrates that the two columns are each frequency synchronized but that the synchronized voltages for each column are different. The numbers in the legend denote the coordinates of each horizontal junction. For example the coordinates $1,2$ denote the horizontal junction in the first row (starting from the bottom row) and second column (starting from the left column).}
\end{figure}

In this section we present some preliminary results on synchronization in disordered two-dimensional (2D) arrays in the presence of shortcuts. Geometrically, we can think of a pristine 2D array as a set of $M$ columns, or ``ladders'', each with $N$ plaquettes, grafted together (see Fig.~\ref{fig:2darray}, which depicts $M=2$, $N=5$). It is well known that in such a geometry, phase locking of all the horizontal junctions can occur in a horizontally-biased uniform array (\textit{i.e.} no critical-current disorder) but that a high-degree of neutral stability is exhibited\cite{Wiesenfeld94}. More precisely, in an array with $M$ columns, there will be a zero-valued Floquet exponent with multiplicity $M$. It was shown in Ref.~\cite{Trees99} that \textit{underdamped} arrays in an external magnetic field perpendicular to the plane of the array could lift this degeneracy via the coupling of the junctions to the external magnetic field in the gauge-invariant phase difference. In the presence of critical current disorder, however, numerical simulations of the RSJ equations revealed that frequency synchronization was no longer possible. In fact, each column or ladder in the array would individually synchronize but that sufficient ``inter-ladder'' coupling was not present to entrain the entire set of horizontal junctions\cite{Whan96,Landsberg2000}. This behavior is shown in Fig.~\ref{fig:cluster} by means of a so-called cluster diagram.
  
\begin{figure}
\includegraphics[scale=0.75]{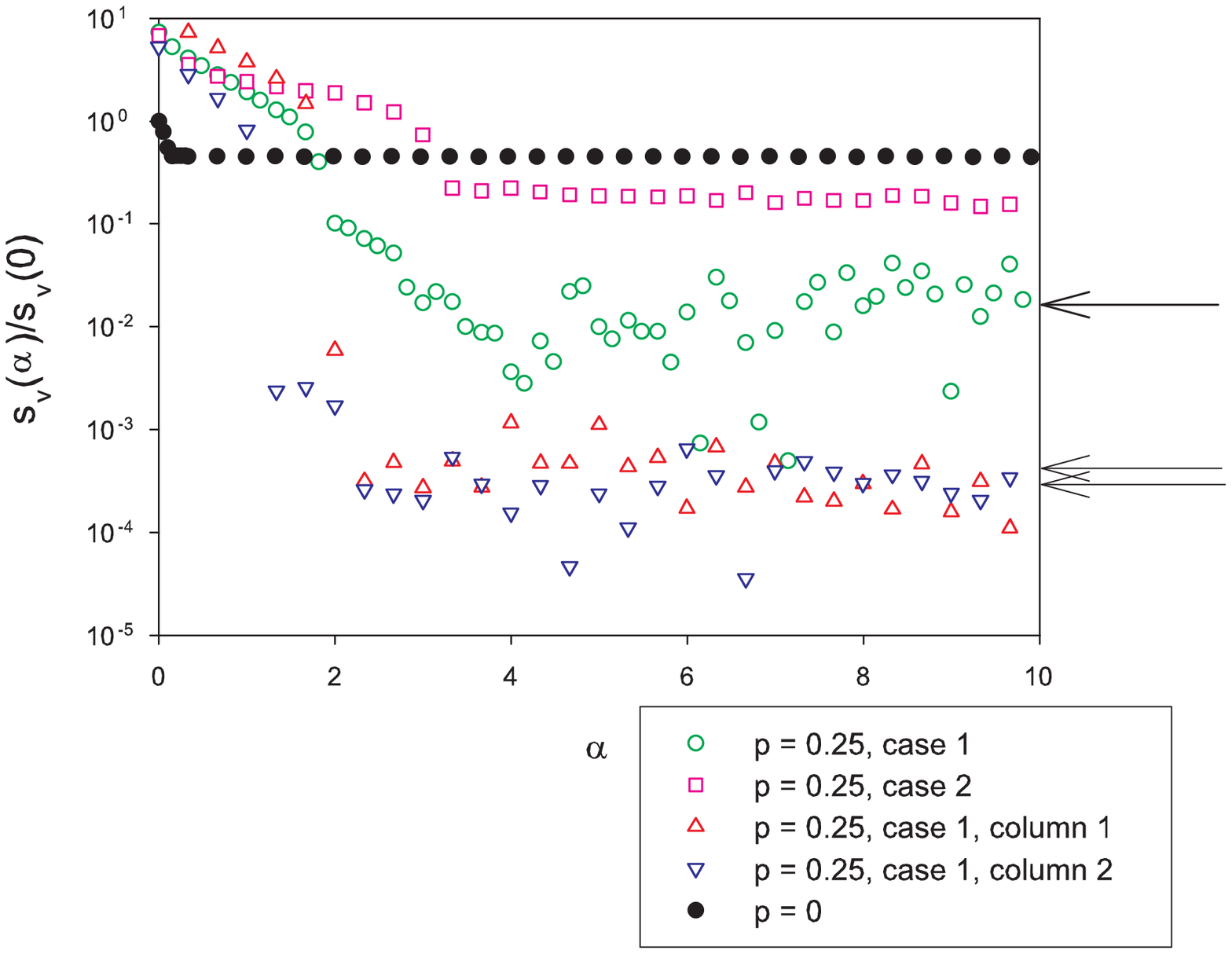}
\caption{\label{fig:52} (Color online) Scaled standard deviation $s_v(\alpha)/s_v(0)$ of time-averaged voltages across the horizontal junctions, plotted versus coupling strength $\alpha$ for an array with $M=2$ columns and $N=5$ plaquettes per column. The bias current is $i_B=5$, and the horizontal junctions are assigned critical currents randomly according to Eq.~\ref{eq:randomic} with $\Delta=0.05$. We assume periodic boundary conditions in the vertical direction. The solid circles are for the pristine array, $p=0$. The hollow circles and squares represent two different realizations of shortcuts at $p=0.25$ in which the added shortcut junctions are critical-current disordered. The hollow up(down) triangles denote the value of $s_v(\alpha)/s_v(0)$ for only the junctions in column one(two). The arrows pointing to the right axis denote the average values of $s_v(\alpha)/s_v(0)$ over the interval $4 < \alpha < 10$ for the hollow circles and both the up and down triangles.}
\end{figure}

To include shortcut connections in the 2D array, we followed a procedure similar to that described in the previous section. For clarity, let each horizontal junction in the array be described by a pair of coordinates $i,j$, where $i$ denotes the row and $j$ the column in which the junction is positioned. To establish a connection between two particular horizontal junctions, say junctions $i,j$ and $k,l$, that are not already nearest neighbors, we add two new junctions, one connecting the left superconducting island of junction $i,j$ to the left island of junction $k,l$ and the second junction is added between the islands on the right sides of $i,j$ and $k,l$. We also allow the shortcut junctions to be critical-current disordered.  Contrary to the case of individual ladder arrays, the effects of shortcuts on synchronization in 2D arrays is not so easily characterized. Figure~\ref{fig:52} shows the scaled standard deviation of the time-averaged voltages, $s_v(\alpha)/s_v(0)$, for the ten horizontal junctions of an overdamped array with $M=2$ and $N=5$ and for $p=0.25$. Note that the ratio does not approach unity for nonzero $p$ as $\alpha\rightarrow 0$ because of the presence of the disordered shortcut junctions. For reference, $s_v(\alpha)/s_v(0)$ for the pristine array is also shown (solid circles). In this case, entrainment is frustrated in that the ratio settles into a clearly nonzero value as the coupling strength is increased. The hollow circles and squares in Fig.~\ref{fig:52} are the values of $s_v(\alpha)/s_v(0)$ for two different realizations of shortcuts at $p=0.25$. For case $2$ in the Figure (hollow squares), the shortcuts have only slightly improved the level of synchronization compared to the pristine case, as $s_v(\alpha)/s_v(0)$ is only reduced by a factor of about $0.35$ compared to its $p=0$ value. Case $1$ (hollow circles) is more interesting in that $s_v(\alpha)/s_v(0)$ is reduced, on average, by an order of magnitude compared to the pristine case by the particular realization of shortcut junctions present. (Note the logarithmic scale on the vertical axis and the topmost arrow on the right axis, which denotes the average value of $s_v(\alpha)/s_v(0)$ for $4 <\alpha <10$.) The noise evident in the results for case $1$ is probably a finite-size effect, but studies of larger arrays are necessary to be sure.

\begin{figure}
\includegraphics[scale=0.6]{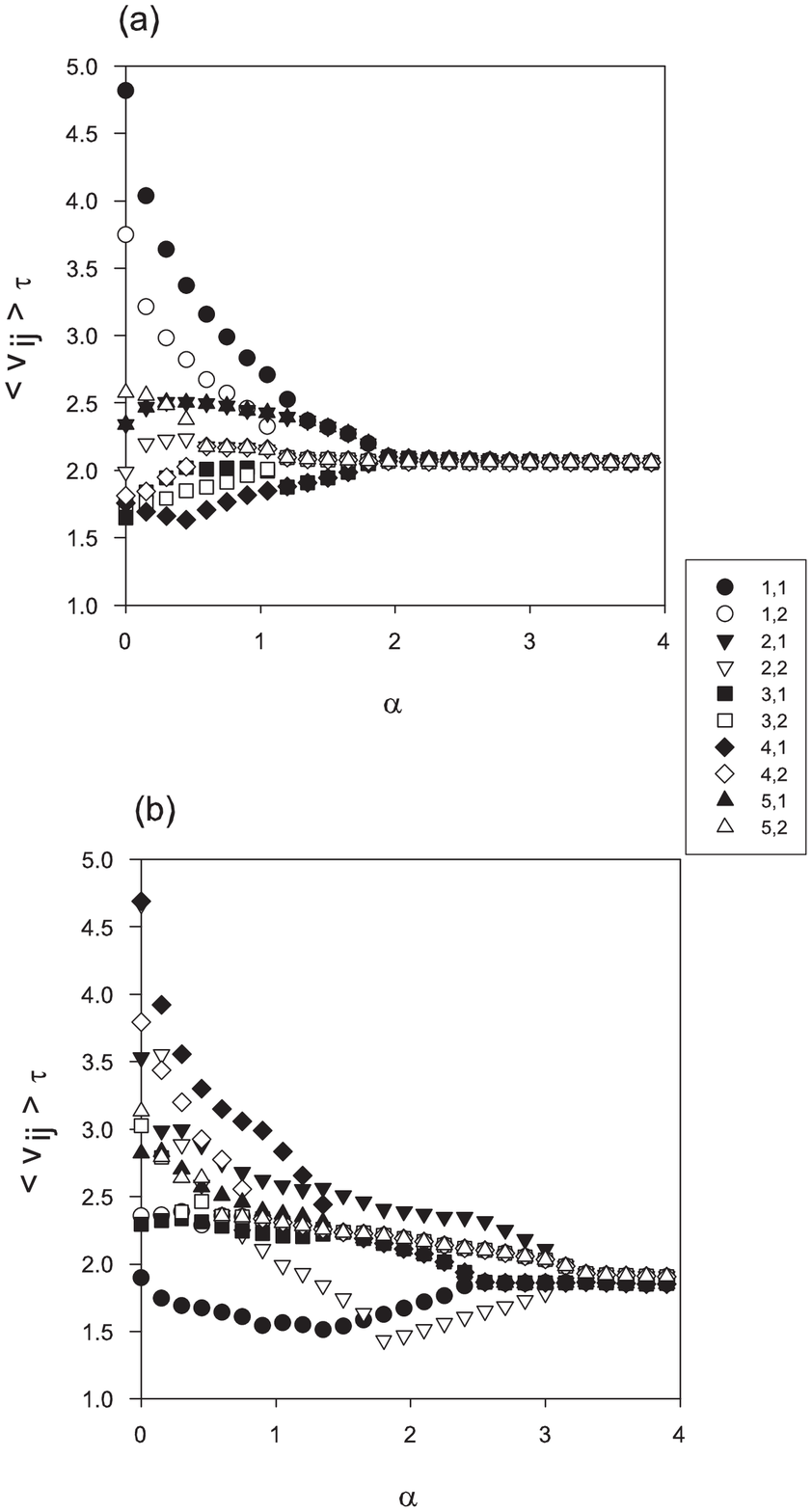}
\caption{\label{fig:cluster52} Cluster diagrams for the time-averaged voltages, $\langle v_{ij}\rangle_{\tau}$, across the horizontal junctions, plotted versus coupling strength $\alpha$ for an array with $M=2$ columns and $N=5$ plaquettes per column. The bias current is $i_B=5$, applied and extracted along each row, and the horizontal junctions are assigned critical currents randomly according to Eq.~\ref{eq:randomic} with $\Delta =0.05$. We assume periodic boundary conditions in the vertical direction. The legend provides the coordinates for each horizontal junction as in Fig.~\ref{fig:cluster}. (a) $p=0.25$, first shortcut realization. (b) $p=0.25$, second shortcut realization.}
\end{figure}

Although array synchronization has clearly not occurred in the second realization of shortcuts in Fig.~\ref{fig:52}, the reduced value of $s_v(\alpha)/s_v(0)$ for case $1$ does not automatically imply entrainment has occurred in that case.  Included in Fig.~\ref{fig:52} are the values of $s_v(\alpha)/s_v(0)$ for the junctions in each column \textit{separately} (hollow triangles). The low average value of these quantities (see the two lower arrows along the right axis) shows that the junctions in a given column are much more strongly entrained to each other than to junctions in the neighboring column.  Thus, the hollow circles in Fig.~\ref{fig:52} correspond, at best, to \textit{weak} intercolumn synchronization. Figures~\ref{fig:cluster52}(a) and~(b) are cluster diagrams for cases $1$ and $2$, respectively, in which the vertical axes of the two plots have the same scale.  Simple visual inspection of the plots suggest that the array is weakly frequency synchronized in case $1$ for $\alpha \agt 2$ but not in case $2$. In fact, we have considered ten different realizations of shortcuts at $p=0.25$ and find this weak synchronization behavior only for one realization (\textit{i.e.} case $1$). The nine remaining realizations resulted in an array that was clearly not entrained, as in case $2$. Based on these results, we thus conclude that shortcuts in 2D arrays, biased in the standard way shown in Fig.~\ref{fig:2darray}, do not significantly enhance the array's ability to synchronize. We see similar effects for $p=0.5$. 

\begin{figure}
\includegraphics[scale=0.6]{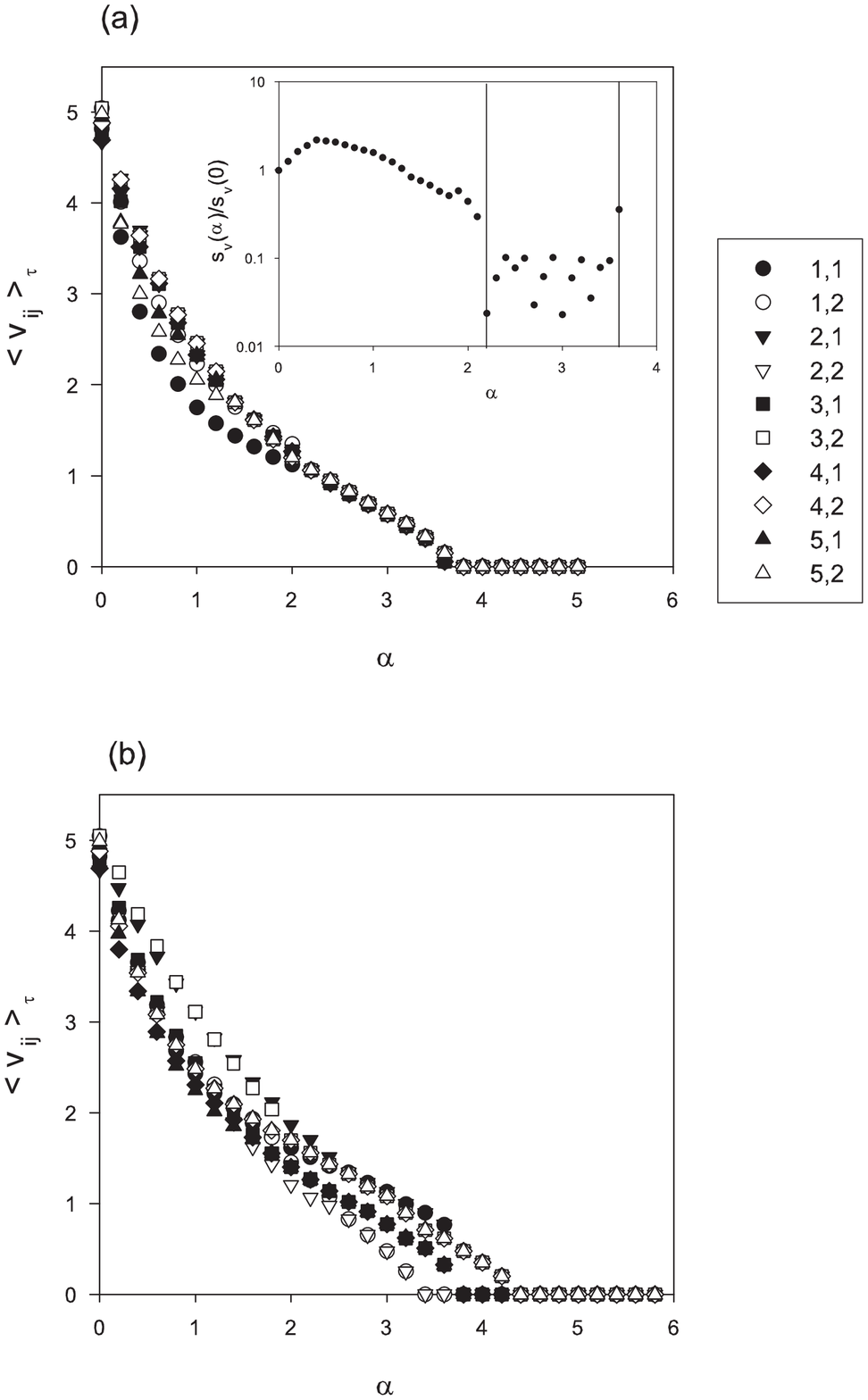}
\caption{\label{fig:SCuniform} Cluster diagrams for the time averaged-voltages, $\langle v_{ij}\rangle_{\tau}$, across the horizontal junctions, plotted versus coupling strength $\alpha$ for an array with $M=2$ and $N=5$. The bias current is $i_B=5$, and the horizontal junctions are assigned critical currents randomly according to Eq.~\ref{eq:randomic} with $\Delta=0.05$. We assume periodic boundary conditions in the vertical direction. All shortcut junctions are identical to the uniform vertical junctions in the pristine array, \textit{i.e.} the shortcuts junctions are not disordered. The legend provides the coordinates for each horizontal junction as in Fig.~\ref{fig:cluster}. (a) $p=0.25$, first shortcut realization. For $2.2\alt\alpha\alt 3.6$, the horizontal junctions are weakly entrained, and for $\alpha\agt 3.6$ the array is in the zero voltage state. The inset shows $s_v(\alpha)/s_v(0)$ versus $\alpha$. The range of $\alpha$ values between the two vertical lines denote the region of weak entrainment. (b) $p=0.25$, second shortcut realization. There is no evidence of entrainment up to $\alpha \approx 4.4$, beyond which the array is in the zero voltage state.}
\end{figure}

We have also considered the effects of \textit{uniform} shortcuts in the 2D array, where each additional shortcut junction is identical to the uniform vertical junctions in the pristine array. As in the case of disordered shortcuts, when we consider ten different realizations at $p=0.25$ we find that in some cases (roughly $40\%$ of the realizations) the array weakly frequency synchronizes and in the remaining cases it clearly does not. Results representative of these two outcomes are shown in Fig.~\ref{fig:SCuniform}. They show an interesting distinction between this case and the case of disordered shortcuts discussed previously: for sufficiently large coupling $\alpha$, all average voltages across the horizontal junctions now go to zero. This behavior is due to the fact that $\alpha=I_{co}/\langle I_c\rangle$ is the ratio of the critical current of the vertical junctions, \textit{now including shortcut junctions}, to the average critical current of the horizontal junctions. As this ratio increases, for a given number and configuration of shortcuts, a value of $\alpha$ is eventually reached for which all the bias current is able to traverse the circuit without exceeding any particular junction's critical current. In other words, the array is biased below its effective critical current in the presence of shortcuts and thus there is a zero average voltage across the array. 

\begin{figure}
\includegraphics[scale=0.75]{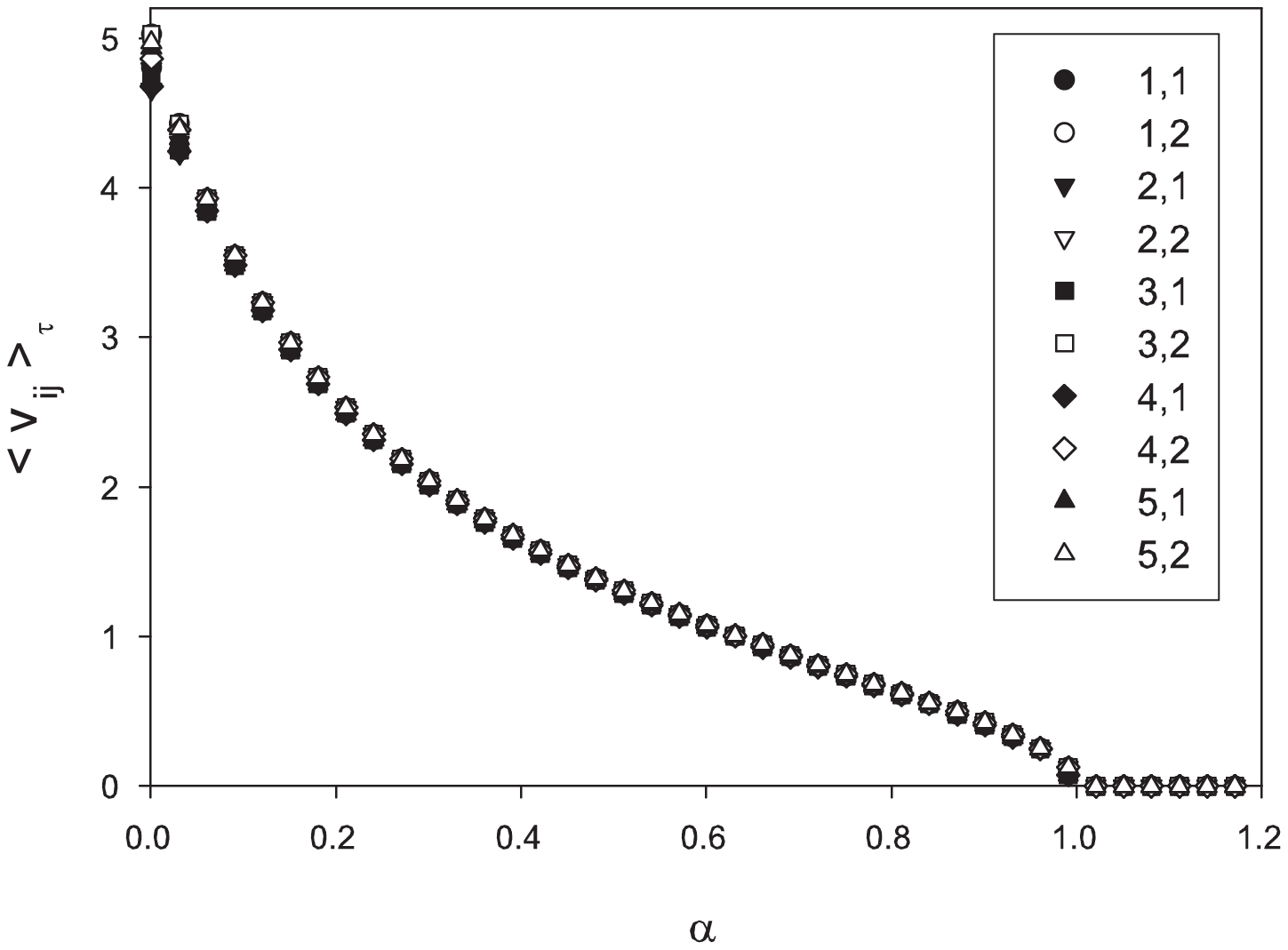}
\caption{\label{fig:p=1} Cluster diagram for the time averaged-voltages, $\langle v_{ij}\rangle_{\tau}$, across the horizontal junctions, plotted versus coupling strength $\alpha$ for an array with $M=2$ and $N=5$. The bias current is $i_B=5$, and the horizontal junctions are assigned critical currents randomly according to Eq.~\ref{eq:randomic} with $\Delta=0.05$. We assume periodic boundary conditions in the vertical direction. All shortcut junctions are identical to the uniform vertical junctions in the pristine array, \textit{i.e.} the shortcuts junctions are not disordered. The legend provides the coordinates for each horizontal junction as in Fig.~\ref{fig:cluster}. Shortcut junctions are added according to $p=1$.}
\end{figure}

The limiting case of $p=1$ means that each horizontal junction is connected to all the remaining horizontal junctions. For this case and with uniform shortcut junctions, we find that the array behaves similarly to Fig.~\ref{fig:SCuniform}(a): there is a range of coupling strengths $\alpha$ for which there is weak entrainment, but for large $\alpha$ the array is in the zero-voltage state (see Fig.~\ref{fig:p=1}).

\section{\label{sec:conclusion}Conclusions}

In this paper we have obtained two main sets of results. First, using a multiple time scale method, we have mapped the exact RCSJ equations for an underdamped ladder with periodic boundary conditions to a second order, locally-coupled Kuramoto model (LKM2). Secondly, we have studied the effects of small world connections on the ability of both ladder and 2D arrays to synchronize. The mapping to the LKM2 is itself useful for two main reasons. First, the synchronization behavior of the Kuramoto model and its variations has been well studied in its own right and could thus shed some light on behavior of actual JJ arrays. Secondly, the LKM2 is solved more quickly on the computer and is easier to understand intuitively than the RCSJ equations. Future work in this area could include using the results of this mapping for the underdamped ladder (as well as the first-order LKM for the overdamped ladder) to arrive at a phase model for the 2D array, as has been suggested in Ref.~\cite{Barbara2002}. Such a phase model for 2D arrays may shed light on why it is difficult for 2D arrays to synchronize.

We have also shown that small-world connections enhance a ladder array's ability to synchronize. This result is not surprising in light of our mapping to the LKM and earlier studies in which the LKM was found to exhibit a mean-field like phase synchronization transition in the presence of shortcuts\cite{Hong2002a}. But we find that SW connections only marginally increase the ability of a 2D array to synchronize. Specifically, for several representative 2D small-world networks, we found only some small fraction of these networks had slightly improved frequency synchronization of the horizontal junctions. In the pristine 2D array ($p=0$) it is well known that no such synchronization, weak or otherwise, is observed over the entire array- hence our characterization that shortcuts are only marginally effective at producing a synchronized array. This conclusion holds whether the additional shortcut junctions are disordered or uniform. Future work in this area could include looking at a broader range of $p$ values as well as looking at larger 2D arrays. It is tempting, but an oversimplification, to think of 2D arrays as mere assemblages of ladder arrays. One source of this temptation is the intriguing fact that the pristine 2D array of disordered junctions will form synchronized clusters consisting of individual ladders. Another is our result that shortcuts can augment synchronization in individual ladders but not in 2D arrays. If one can produce a mapping, even approximate, of the RCSJ equations for the 2D array to a phase model of the Winfree type, this could be very helpful in understanding the rich and perplexing dynamical behavior of 2D arrays. 

\begin{acknowledgments}
BRT wishes to thank the hospitality of the physics department at The Ohio State University and Ohio Wesleyan University for support. VS acknowledges Ohio Wesleyan for support for summer research. DS acknowledges the Ohio Supercomputing Center for a grant of time and NSF grant DMR01-04987 for support.
\end{acknowledgments}

\appendix*
\section{\label{app:details}Multiple Time Scale Analysis}
We substitute Eq.~\ref{eq:expansion} into Eq.~\ref{eq:scaledstart}. To help organize the terms according to the order in $\epsilon$, we also write the nonlinear terms as follows (as in Ref.~\cite{Watanabe97}):
\begin{subequations}
\begin{eqnarray}
\sin\gamma_j& =&\sum_{n=0}^{\infty}\epsilon^n S_{n,j} \label{eq:S} \\
\sin\left(\frac{\gamma_{j+\delta}-\gamma_j}{2}\right)&=&\sum_{n=0}^{\infty}\epsilon^n R_{n,j+\delta}, \label{eq:R}
\end{eqnarray}
\end{subequations}
where $S_{0,j}=\sin\gamma_{0,j}$, $S_{1,j}=\gamma_{1,j}\cos\gamma_{0,j}$, $S_{2,j}=\gamma_{2,j}\cos\gamma_{0,j}-\frac{1}{2}\gamma^{2}_{1,j}\sin\gamma_{0,j}$, $R_{0,j+\delta}=\sin\left[(\gamma_{0,j+\delta}-\gamma_{0,j})/2\right]$, $R_{1,j+\delta}=\frac{1}{2}\cos\left[(\gamma_{0,j+\delta}-\gamma_{0,j})/2\right](\gamma_{1,j+\delta}-\gamma_{1,j})$, and $R_{2,j+\delta}=\frac{1}{2}\cos\left[(\gamma_{0,j+\delta}-\gamma_{0,j})/2\right](\gamma_{2,j+\delta}-\gamma_{2,j})-\frac{1}{8}\sin\left[(\gamma_{0,j+\delta}-\gamma_{0,j})/2\right](\gamma_{1,j+\delta}-\gamma_{1,j})^2$. So Eq.~\ref{eq:scaledstart} can then be written as
\begin{eqnarray}
1=i_{cj}\tilde{\beta_c}\sum_{n=0}^{\infty}\epsilon^n\left[\partial_{0}^{2} + 2\epsilon\partial_0\partial_1 + \epsilon^2\left(2\partial_0\partial_2+\partial^{2}_{1}\right) + 2\epsilon^3\left(\partial_0\partial_3+\partial_1\partial_2\right)\right]\gamma_{n,j} +\nonumber\\
i_{cj}\sum_{n=0}^{\infty}\epsilon^n\left[\partial_0 + \epsilon\partial_1 + \epsilon^2\partial_2 + \epsilon^3\partial_3\right]\gamma_{n,j}+\epsilon i_{cj}\sum_{n=0}^{\infty}\epsilon^nS_{n,j}-\epsilon\alpha\sum_{n=0}^{\infty}\sum_{\delta=\pm 1} \epsilon^nR_{n,j+\delta}.
\label{eq:A3}
\end{eqnarray}
Extracting all terms of ${\cal O}(\epsilon^0)$ yields
\begin{equation}
1=i_{cj}\tilde{\beta_c}\partial^{2}_{0}\gamma_{0,j}+i_{cj}\partial_0\gamma_{0,j}.
\label{eq:orderzero}
\end{equation}
The solution to the homogeneous version of Eq.~\ref{eq:orderzero} is $\gamma_{0,j}=A+Be^{-T_0/\tilde{\beta_c}}$, where $A$ and $B$ are constants \textit{with respect to the $T_0$ time scale}. The exponential term is dropped because it represents transient behavior. We take a particular solution to the inhomogeneous equation of the form $\gamma_{0,j}^{\mbox{(p)}}=C_{j}^{(0)}T_0$, where $C_{j}^{(0)}$ is independent of $T_0$. Substitution of $\gamma_{0}^{\mbox{(p)}}$ into Eq.~\ref{eq:orderzero} yields $C_{j}^{(0)}=1/i_{cj}$. So the solution to Eq.~\ref{eq:orderzero} can be written as
\begin{equation}
\gamma_{0,j}=\frac{T_0}{i_{cj}}+\phi_j(T_1,T_2,T_3),
\label{eq:gammazero}
\end{equation}
where one can think of $\phi_j$ as describing the slow phase dynamics of $\gamma_{0,j}$.

Setting the coefficients of the ${\cal O}(\epsilon^1)$ terms in Eq.~\ref{eq:A3} to zero gives:
\begin{equation}
0=\tilde{\beta_c}\left[\partial^{2}_{0}\gamma_{1,j} + 2\partial_0\partial_1\gamma_{0,j}\right]+\left[\partial_0\gamma_{1,j}+\partial_1\gamma_{0,j}\right] + S_{0,j}-\frac{\alpha}{i_{cj}}\sum_{\delta=\pm 1}R_{0,j+\delta},
\label{eq:orderone}
\end{equation}
where we have divided by a factor of $i_{cj}$. This result is written as
\begin{equation}
\tilde{\beta_c}\partial^{2}_{0}\gamma_{1,j}=-2\tilde{\beta_c}\partial_0\partial_1\gamma_{0,j}-\partial_1\gamma_{0,j}-S_{0,j}+\frac{\alpha}{i_{cj}}\sum_{\delta=\pm 1}R_{o,j+\delta}.
\label{eq:orderoneb}
\end{equation}
Based on Eq.~\ref{eq:gammazero} we can calculate the derivatives on the right side of Eq.~\ref{eq:orderoneb}. We find
\[
\partial_0\partial_1\gamma_{0,j}=\frac{\partial}{\partial T_0}\frac{\partial}{\partial T_1}\left[\frac{T_0}{i_{cj}} + \phi_j(T_1,T_2,T_3)\right] = \frac{\partial}{\partial T_0}\left(\frac{\partial \phi_j}{\partial T_1}\right)=0,
\]
because $\phi_j$ is independent of $T_0$. Also
\[
\partial_1\gamma_{0,j}=\frac{\partial}{\partial T_1}\left[ \frac{T_0}{i_{cj}}+\phi_j\right] = \frac{\partial\phi_j}{\partial T_1}=\partial_1\phi_j.
\]
Next we look at the nonlinear terms on the right side of Eq.~\ref{eq:orderoneb}:
\begin{eqnarray*}
S_{0,j}&=&\sin\left(\frac{T_0}{i_{cj}}+\phi_j\right), \nonumber \\
R_{0,j+\delta}&=&\sin\left[\frac{\left(\frac{T_0}{i_{cj}}-\frac{T_0}{i_{c,j+\delta}}\right) + \left(\phi_{j+\delta}-\phi_j\right)}{2}\right]\approx\sin\left[\frac{\phi_{j+\delta}-\phi_j}{2}\right], \nonumber \\
\end{eqnarray*}
where in the limit of small disorder we have assumed $T_0(1/i_{cj} - 1/i_{c,j+\delta})\approx 0$. So Eq.~\ref{eq:orderoneb} can be written as
\begin{equation}
\tilde{\beta_c}\partial_0\gamma_{1,j}+\partial_0\gamma_{1,j}=-\partial_1\phi_j -\sin\left(\frac{T_0}{i_{cj}}+\phi_j\right) + M_j,
\label{eq:orderonec}
\end{equation}
where $M_j=\frac{\alpha}{i_{cj}}\sum_{\delta}\sin\left[(\phi_{j+\delta}-\phi_j)/2\right]$ is a constant with respect to $T_0$. Temporarily ignoring the derivative $\partial_1\phi_j$ on the right side of Eq.~\ref{eq:orderonec} and using the trigonometric identity for the sine of a sum of two quantities, we find a particular solution to Eq.~\ref{eq:orderonec} of the form
\begin{equation}
\gamma_{1,j}=M_jT_0+ C_{j}^{(1)}\sin\left(\frac{T_0}{i_{cj}}\right) + D_{j}^{(1)}\cos\left(\frac{T_0}{i_{cj}}\right),
\label{eq:gammaone}
\end{equation}
where 
\begin{eqnarray}
C_{j}^{(1)}(T_1,T_2)=\frac{i_{cj}^{2}\left(\tilde{\beta_c}\cos\phi_j - i_{cj}\sin\phi_j\right)}{i_{cj}^{2}+\tilde{\beta_c}^2}, \label{eq:C} \\
D_{j}^{(1)}(T_1,T_2)=\frac{i_{cj}^{2}\left(\tilde{\beta_c}\sin\phi_j + i_{cj}\cos\phi_j\right)}{i_{cj}^{2}+\tilde{\beta_c}^2}. \label{eq:D}
\label{eq:CandD}
\end{eqnarray}
The term $M_jT_0$ in Eq.~\ref{eq:gammaone} represents a secular term that grows without bound as $T_0\rightarrow\infty$. To remove this term from the solution we impose the condition
\[
-\partial_1\phi_j + M_j =0,
\]
which gives
\begin{equation}
\frac{\partial\phi_j}{\partial T_1}=\frac{\alpha}{i_{cj}}\sum_{\delta =\pm 1}\sin\left[\frac{\phi_{j+\delta}-\phi_j}{2}\right].
\label{eq:dT1}
\end{equation}
This in turn gives a solution for $\gamma_{1,j}$ that is Eq.~\ref{eq:gammaone} without the secular term. Note that Eq.~\ref{eq:dT1} measures the rate of change of $\phi_j$, and hence $\gamma_{0,j}$, with respect to the slow time scale $T_1$.

Next, we look at all terms in Eq.~\ref{eq:A3} that are ${\cal O}(\epsilon^2)$:
\begin{equation}
\tilde{\beta_c}\partial_{0}^{2}\gamma_{2,j}+\partial_0\gamma_{2,j}=-2\tilde{\beta_c}\partial_0\partial_1\gamma_{1,j}-\tilde{\beta_c}\left(2\partial_0\partial_2+\partial_{1}^{2}\right)\gamma_{0,j}-\partial_1\gamma_{1,j}-\partial_2\gamma_{0,j}-S_{1,j}+\frac{\alpha}{i_{cj}}\sum_{\delta=\pm 1}T_{1,j+\delta}.
\label{eq:ordertwo}
\end{equation}
Using the known results for $\gamma_{0,j}$ and $\gamma_{1,j}$ to calculate the derivatives on the right side of Eq.~\ref{eq:ordertwo}, and using the expression for $S_{1,j}$ and $T_{1,j+\delta}$, means that Eq.~\ref{eq:ordertwo} can be written (after some algebra) as
\begin{equation}
\tilde{\beta_c}\partial_{0}^{2}\gamma_{2,j}+\partial_0\gamma_{2,j}=-\partial_2\phi_j + V_j +W_j\sin\left(\frac{T_0}{i_{cj}}\right)+X_j\cos\left(\frac{T_0}{i_{cj}}\right) + Y_j\sin\left(\frac{2T_0}{i_{cj}}\right)+Z_j\cos\left(\frac{2T_0}{i_{cj}}\right),
\label{eq:ordertwob}
\end{equation}
where
\[
V_j=-\tilde{\beta_c}\partial_{1}^{2}\phi_j - \frac{i_{cj}^{3}}{2\left(i_{cj}^{2}+\tilde{\beta_c}^2\right)},
\]
\[
W_j=\frac{2\tilde{\beta_c}}{i_{cj}}C_{j}^{(1)}\partial_1\phi_j + D_{j}^{(1)}\partial_1\phi_j+\frac{\alpha}{2i_{cj}}\sum_{\delta=\pm 1}\left(C_{j+\delta}^{(1)}-C_{j}^{(1)}\right)\cos\left[\frac{\phi_{j+\delta}-\phi_j}{2}\right],
\]
\[
X_j=\frac{2\tilde{\beta_c}}{i_{cj}}D_{j}^{(1)}\partial_1\phi_j + C_{j}^{(1)}\partial_1\phi_j+\frac{\alpha}{2i_{cj}}\sum_{\delta=\pm 1}\left(D_{j+\delta}^{(1)}-D_{j}^{(1)}\right)\cos\left[\frac{\phi_{j+\delta}-\phi_j}{2}\right],
\]
\[
Y_j=-\frac{1}{2}\left[C_{j}^{(1)}\cos\phi_j-D_{j}^{(1)}\sin\phi_j\right],
\]
\[
Z_j=-\frac{1}{2}\left[C_{j}^{(1)}\sin\phi_j+D_{j}^{(1)}\cos\phi_j\right],
\]
where $C_{j}^{(1)}$ and $D_{j}^{(1)}$ are given by Eqs.~\ref{eq:C} and~\ref{eq:D}.
As with the first order case, we want a solution to Eq.~\ref{eq:ordertwob} that does not have any secular terms. Therefore we must impose the condition
\begin{equation}
\frac{\partial\phi_j}{\partial T_2}=V_j = -\tilde{\beta_c}\frac{\partial^2\phi_j}{\partial T_{1}^{2}}-\frac{i_{cj}^{3}}{2\left(i_{cj}^{2}+\tilde{\beta_c}^2\right)}.
\label{eq:dT2}
\end{equation}
Based on Eq.~\ref{eq:dT1} it is possible to calculate $\partial^2\phi_j/\partial T_ {1}^{2}$, which appears on the right side of Eq.~\ref{eq:dT2}. We find
\begin{eqnarray}
\frac{\partial^2\phi_j}{\partial T_{1}^{2}}&=&\frac{\alpha}{i_{cj}}\frac{\partial}{\partial T_1}\sum_{\delta=\pm 1}\sin\left[\frac{\phi_{j+\delta}-\phi_j}{2}\right] \nonumber\\
& = & \frac{\alpha^2}{2 i_{cj}^{2}}\left[\sum_{\delta=\pm 1}\sin\left(\phi_j-\phi_{j+\delta}\right) + \frac{1}{2}\sum_{\delta =\pm 2}\sin\left(\frac{\phi_{j+\delta}-\phi_j}{2}\right)-\sin^2\left(\frac{\nabla^2\phi_j}{2}\right) + \frac{1}{2}\sum_{\delta=\pm 1}\sin\left(\frac{\nabla^2\phi_{j+\delta}}{2}\right)\right] \nonumber \\
 &\equiv & \frac{\alpha^2}{2i_{cj}^2}{\cal Z}_j, \label{eq:seconddT1}
\end{eqnarray}
where we approximated $1/(i_{cj}i_{c,j\pm 1})$ with $1/i_{cj}^{2}$. Note that $\nabla^2\phi_j\equiv \phi_{j+1}-2\phi_j+\phi_{j-1}$. Substituting Eq.~\ref{eq:seconddT1} into Eq.~\ref{eq:dT2} yields
\begin{equation}
\frac{\partial\phi_j}{\partial T_2}=-\frac{i_{cj}^{3}}{2\left(i_{cj}^{2}+\tilde{\beta_c}^2\right)}-\frac{\alpha^2\tilde{\beta_c}}{2i_{cj}^{2}}{\cal Z}_j.
\label{eq:dT2final}
\end{equation}

Our next step is to calculate the derivative $d\gamma_{0,j}/d\tau$ with respect to the original dimensionless time variable $\tau=\tilde{\tau}/i_B$. We find
\begin{eqnarray}
\frac{d\gamma_{0,j}}{d\tau}&=&i_B\frac{d\gamma_{0,j}}{d\tilde{\tau}}=i_B\left[\partial_0 +\epsilon\partial_1 + \epsilon^2\partial_2\right]\left[\frac{T_0}{i_{cj}}+\phi_j(T_1,T_2)\right] \nonumber \\
 &=& \frac{i_B}{i_{cj}}+\frac{\alpha}{i_{cj}}\sum_{\delta=\pm 1}\sin\left[\frac{\phi_{j+\delta}-\phi_j}{2}\right]-\frac{i_{cj}^{3}/i_B}{2\left(i_{cj}^{2}+i_{B}^{2}\beta_c^2\right)}-\frac{\alpha^2\beta_c}{2i_{cj}^{2}}{\cal Z}_j, \label{eq:dgdtau}
\end{eqnarray}
where use was made of the expressions $\epsilon=1/i_B$ and $\tilde{\beta_c}=i_B\beta_c$. It is convenient to define the quantity
\begin{equation}
\Omega_j=\frac{i_B}{i_{cj}}\left[1 - \frac{\left(i_{cj}/i_B\right)^4}{2\left\{\beta_{c}^{2}+\left(i_{cj}/i_B\right)^2\right\}}\right].
\label{eq:omegaj}
\end{equation}
Physically, one can think of $\Omega_j$ as the angular frequency(average voltage) of oscillator(junction) $j$ in the absence of coupling. Then Eq.~\ref{eq:dgdtau} can be written as
\[
\frac{d\gamma_{0,j}}{d\tau}=\Omega_j+\frac{\alpha}{i_{cj}}\sum_{\delta=\pm 1}\sin\left[\frac{\phi_{j+\delta}-\phi_j}{2}\right]-\frac{\alpha^2\beta_c}{2i_{cj}^{2}}{\cal Z}_j.
\]
It is also useful to calculate the second derivative of $\gamma_{0,j}$:
\begin{eqnarray}
\frac{d^2\gamma_{0,j}}{d\tau^2}&=&i_{B}^{2}\frac{d^2\gamma_{0,j}}{d\tilde{\tau}^2}=i_{B}^{2}\left[\partial_{0}^{2}+2\epsilon\partial_0\partial_1+\epsilon^2\left(2\partial_0\partial_1+\partial_1^2\right)\right]\phi_j \nonumber \\
 & = & \frac{\alpha^2}{2i_{cj}^{2}}{\cal Z}_j, \label{eq:dgdtau2}
\end{eqnarray}
where we made use of Eq.~\ref{eq:seconddT1}. It is common practice at this juncture to replace the symbol $\gamma_{0,j}$ in Eq.~\ref{eq:dgdtau} with $\phi_j$, which we shall also do in Eq.~\ref{eq:dgdtau2}.

Finally, motivated by the structure of Eq.~\ref{eq:start}, consider the combination of terms 
\[
i_{cj}\beta_c\frac{d^2\phi_j}{d\tau^2}+i_{cj}\frac{d\phi_j}{d\tau}.
\]
Substituting for the derivatives from Eq.~\ref{eq:dgdtau} and~\ref{eq:dgdtau2} and dividing through by a factor of $i_{cj}$ results in the expression
\begin{equation}
\beta_c\frac{d^2\phi_j}{d\tau^2}+\frac{d\phi_j}{d\tau}=\Omega_j +\frac{\alpha}{i_{cj}}\sum_{\delta=\pm 1}\sin\left[\frac{\phi_{j+\delta}-\phi_j}{2}\right],
\label{eq:big1}
\end{equation}
which is Eq.~\ref{eq:workingresult} in Sec.~\ref{sec:multipletime}.
A straightforward but tedious continuation of the analysis to ${\cal O}(\epsilon^3)$ then leads to Eq.~\ref{eq:fullresult} in Sec.~\ref{sec:multipletime}.


\begin{thebibliography}{99}
\bibitem{comment1} We also use fluxoid quantization to gain one more constraint, $\sum_{j=1}^{N}\psi_{1,j}=0$. This leaves us with $2N-1$ unknowns which is the same number of unknowns obtained if we use the phases of the wavefunctions describing the superconducting islands as the dynamical variables.
\end{thebibliography}
\end{document}